%% file: k18br.tex
\documentclass[seceq]{ptptex}

\usepackage{graphicx}
\usepackage[dvips]{color}
\usepackage{multirow}

\newcommand{\KEK}{1}
\newcommand{\RCNP}{2}
\newcommand{\Victoria}{3}
\newcommand{\Seoul}{4}
\newcommand{\INFNHH}{5}
\newcommand{\SMI}{6}
\newcommand{\Torino}{7}
\newcommand{\TorinoU}{8}
\newcommand{\Frascati}{9}
\newcommand{\Osaka}{10}
\newcommand{\Kyoto}{11}
\newcommand{\Tokyo}{12}
\newcommand{\OsakaE}{13}
\newcommand{\RIKEN}{14}
\newcommand{\TITEC}{15}
\newcommand{\TUM}{16}
\newcommand{\komaba}{17}
\newcommand{\Tohoku}{18}
\newcommand{\ECUTUM}{19}
\newcommand{\KIRAMS}{20}

\title{
The K1.8BR spectrometer system at J-PARC
}

\author{
Keizo \textsc{Agari}$^{\KEK}$,
Shuhei \textsc{Ajimura}$^{\RCNP}$,
George \textsc{Beer}$^{\Victoria}$,
Hyoungchan \textsc{Bhang}$^{\Seoul}$,
Mario \textsc{Bragadireanu}$^{\INFNHH}$,
Paul \textsc{Buehler}$^{\SMI}$,
Luigi \textsc{Busso}$^{\Torino,\TorinoU}$,
Michael \textsc{Cargnelli}$^{\SMI}$,
Seonho \textsc{Choi}$^{\Seoul}$,
Catalina \textsc{Curceanu}$^{\Frascati}$,
Shun \textsc{Enomoto}$^{\Osaka}$,
Diego \textsc{Faso}$^{\Torino,\TorinoU}$,
Hiroyuki \textsc{Fujioka}$^{\Kyoto}$,
Yuya \textsc{Fujiwara}$^{\Tokyo}$,
Tomokazu \textsc{Fukuda}$^{\OsakaE}$,
Carlo \textsc{Guaraldo}$^{\Frascati}$,
Tadashi \textsc{Hashimoto}$^{\Tokyo}$,
Ryugo S. \textsc{Hayano}$^{\Tokyo}$,
Toshihiko \textsc{Hiraiwa}$^{\Kyoto}$,
Erina \textsc{Hirose}$^{\KEK}$,
Masaharu \textsc{Ieiri}$^{\KEK}$,
Masami \textsc{Iio}$^{\KEK}$,
Mihai  \textsc{Iliescu}$^{\Frascati}$,
Kentaro \textsc{Inoue}$^{\Osaka}$,
Yosuke \textsc{Ishiguro}$^{\Kyoto}$,
Takashi \textsc{Ishikawa}$^{\Tokyo}$,
Shigeru \textsc{Ishimoto}$^{\KEK}$,
Tomoichi \textsc{Ishiwatari}$^{\SMI}$,
Kenta \textsc{Itahashi}$^{\RIKEN}$,
Masaaki \textsc{Iwai}$^{\KEK}$,
Masahiko \textsc{Iwasaki}$^{\TITEC,\RIKEN}$,
Yutaka \textsc{Kakiguchi}$^{\KEK}$,
Yohji \textsc{Katoh}$^{\KEK}$,
Shingo \textsc{Kawasaki}$^{\Osaka}$,
Paul \textsc{Kienle}$^{\TUM}$,
Hiroshi \textsc{Kou}$^{\TITEC}$,
Yue \textsc{Ma}$^{\RIKEN}$,
Johann \textsc{Marton}$^{\SMI}$,
Yasuyuki \textsc{Matsuda}$^{\komaba}$,
Michifumi \textsc{Minakawa}$^{\KEK}$,
Yutaka \textsc{Mizoi}$^{\OsakaE}$,
Ombretta \textsc{Morra}$^{\Torino}$,
Ryotaro \textsc{Muto}$^{\KEK}$,
Tomofumi \textsc{Nagae}$^{\Kyoto}$,
Megumi \textsc{Naruki}$^{\KEK}$,
Hiroyuki \textsc{Noumi}$^{\RCNP}$,
Hiroaki \textsc{Ohnishi}$^{\RIKEN}$,
Shinji \textsc{Okada}$^{\RIKEN}$,
Haruhiko \textsc{Outa}$^{\RIKEN}$,
Kristian \textsc{Piscicchia}$^{\Frascati}$,
Marco \textsc{Poli Lener}$^{\Frascati}$,
Antonio \textsc{Romero Vidal}$^{\Frascati}$,
Yuta \textsc{Sada}$^{\Kyoto}$,
Atsushi \textsc{Sakaguchi}$^{\Osaka}$,
Fuminori \textsc{Sakuma}$^{\RIKEN}$\footnote{E-mail: sakuma@ribf.riken.jp},
Masaharu \textsc{Sato}$^{\Tokyo}$,
Yoshinori \textsc{Sato}$^{\KEK}$,
Shin'ya \textsc{Sawada}$^{\KEK}$,
Alessandro \textsc{Scordo}$^{\Frascati}$,
Michiko \textsc{Sekimoto}$^{\KEK}$,
Hexi \textsc{Shi}$^{\Tokyo}$,
Yoshihisa \textsc{Shirakabe}$^{\KEK}$,
Diana \textsc{Sirghi}$^{\Frascati,\INFNHH}$,
Florin \textsc{Sirghi}$^{\Frascati,\INFNHH}$,
Ken \textsc{Suzuki}$^{\SMI}$,
Shoji \textsc{Suzuki}$^{\KEK}$,
Takatoshi \textsc{Suzuki}$^{\Tokyo}$,
Yoshihiro \textsc{Suzuki}$^{\KEK}$,
Hitoshi \textsc{Takahashi}$^{\KEK}$,
Kazuhiro \textsc{Tanaka}$^{\KEK}$,
Nobuaki \textsc{Tanaka}$^{\KEK}$,
Hideyuki \textsc{Tatsuno}$^{\Frascati}$,
Makoto \textsc{Tokuda}$^{\TITEC}$,
Dai \textsc{Tomono}$^{\RIKEN}$,
Akihisa \textsc{Toyoda}$^{\KEK}$,
Kyo \textsc{Tsukada}$^{\Tohoku}$,
Oton \textsc{Vazquez Doce}$^{\Frascati,\ECUTUM}$,
Hiroaki \textsc{Watanabe}$^{\KEK}$,
Eberhard \textsc{Widmann}$^{\SMI}$,
Barbara K. \textsc{W\"{u}nschek}$^{\SMI}$,
Yutaka \textsc{Yamanoi}$^{\KEK}$,
Toshimitsu \textsc{Yamazaki}$^{\Tokyo,\RIKEN}$,
Heejoong \textsc{Yim}$^{\KIRAMS}$,
and Johann \textsc{Zmeskal}$^{\SMI}$
}

\inst{
$^{\KEK}$High Energy Accelerator Research Organization (KEK), Tsukuba,
305-0801, Japan \\
$^{\RCNP}$Research Center for Nuclear Physics (RCNP), Osaka University,
 Osaka, 567-0047, Japan \\
$^{\Victoria}$Department of Physics and Astronomy, University of
Victoria, Victoria BC V8W 3P6, Canada \\
$^{\Seoul}$Department of Physics, Seoul National University, Seoul,
 151-742, South Korea \\
$^{\INFNHH}$National Institute of Physics and Nuclear Engineering -
IFIN HH, Romania \\
$^{\SMI}$Stefan-Meyer-Institut f\"{u}r subatomare Physik, A-1090
Vienna, Austria \\
$^{\Torino}$INFN Sezione di Torino, Torino, Italy \\
$^{\TorinoU}$Dipartimento di Fisica Generale, Universita' di Torino,
Torino, Italy \\
$^{\Frascati}$Laboratori Nazionali di Frascati dell' INFN, I-00044
Frascati, Italy \\
$^{\Osaka}$Department of Physics, Osaka University, Osaka, 560-0043,
Japan \\
$^{\Kyoto}$Department of Physics, Kyoto University, Kyoto, 606-8502,
Japan \\
$^{\Tokyo}$Department of Physics, The University of Tokyo, Tokyo,
113-0033, Japan \\
$^{\OsakaE}$Laboratory of Physics, Osaka Electro-Communication
University, Osaka, 572-8530, Japan \\
$^{\RIKEN}$RIKEN Nishina Center, RIKEN, Wako, 351-0198, Japan \\
$^{\TITEC}$Department of Physics, Tokyo Institute of Technology, Tokyo,
152-8551, Japan \\
$^{\TUM}$Technische Universit\"{a}t M\"{u}nchen, D-85748, Garching,
Germany \\
$^{\komaba}$Graduate School of Arts and Sciences, The University of
Tokyo, Tokyo, 153-8902, Japan \\
$^{\Tohoku}$Department of Physics, Tohoku University, Sendai, 980-8578,
Japan \\
$^{\ECUTUM}$Excellence Cluster Universe, Technische Universit\"{a}t
M\"{u}nchen, D-85748, Garching, Germany \\
$^{\KIRAMS}$Korea Institute of Radiological and Medical Sciences
(KIRAMS), Seoul, 139-706, South Korea
}

\abst{
A new spectrometer system was designed and constructed at the secondary
beam line K1.8BR in the hadron hall of J-PARC to investigate
$\bar K N$ interactions and $\bar K$-nuclear bound systems.
The spectrometer consists of a high precision beam line spectrometer, a
liquid $^3$He/$^4$He/D$_2$ target system, a Cylindrical Detector System
that surrounds the target to detect the decay particles from the target
region, and a neutron time-of-flight counter array located $\sim$15~m
downstream from the target position.
Details of the design, construction, and performance of the detector
components are described.
}

\begin{document}

\maketitle

\section{Introduction}
The $\bar K N$ interaction is one of the keys to understanding meson-baryon
interactions in low energy quantum chromodynamics (QCD) incorporating
three flavors in the nuclear system.
Precise measurements of elementary $\bar K N$ interactions and
investigations of $\bar K$-nuclear bound systems ($\bar K$ nuclei) are
currently hot topics.
The physics goal using the secondary beam line K1.8BR is to focus on a
detailed investigation of the $\bar K N$ interaction and $\bar K$
nuclei.

Extensive measurements of the anti-kaonic hydrogen atom~\cite{K-Helium}
and low-energy $\bar K N$ scattering~\cite{Kp} have shown that the $\bar
K N$ interaction is strongly attractive, but this is still not fully
understood.
In particular, the $\Lambda(1405)$ resonance, which appears 27 MeV below
the $K^-p$ threshold, has been the biggest issue in understanding the $\bar
K N$ interactions and $\bar K$ nuclei structure~\cite{AY,Lambda1405}.
As a consequence of the strongly attractive $I = 0$ $\bar K N$
interaction, the concept of $\Lambda(1405)$ as a ``seed'' for possible $\bar
K$-nuclear quasi-bound states has been widely discussed in recent years.
In addition, the simplest $\bar K$-nuclear cluster, $K^-pp$, is of
special interest because it is the lightest $S = -1$ $\bar K$ nucleus,
from which the evolution of dense $\bar K$ nuclei such as $K^-ppn$ and
$K^-K^-pp$ is naturally expected.
In recent years many theoretical works have supported the existence of
the $K^-pp$ bound state, but the predicted binding energies and widths
are widely divergent~\cite{AY,Knucl}.
Experimentally, however, only a small amount of information is
available~\cite{exp-Knucl}, which is not sufficient to discriminate
between a variety of conflicting interpretations.

Experiments using elementary $\bar K$-induced reactions are expected
to clarify such controversial $\bar K N$ interactions and $\bar K$
nuclei.
At the K1.8BR beam line at J-PARC, the following three new experiments
with the $K^-$ beam have been proposed and approved.

\subsection{A search for deeply bound kaonic nuclear states by the
in-flight $^3$He$(K^-, n)$ reaction (the E15 experiment)}
The E15 experiment searches for the simplest $\bar K$ nucleus,
$K^-pp$~\cite{E15}, whose existence was predicted by the pioneering
work of Y.~Akaishi and T.~Yamazaki\cite{AY}.
The goal of the experiment is to investigate the structure and decay of
$K^-pp$.
If the deeply bound $K^-pp$ state is found to exist as
predicted\cite{AY}, we can extend our experimental study to heavier
nuclei, such as $K^-ppn$, $K^-ppp$, etc., which are also predicted to be
deeply bound and of high density.
Such ultra-high-density matter is the gateway toward kaon-condensed
matter, where the chiral symmetry is expected to be restored.

The experiment aims to identify the nature of the $K^-pp$ bound
state by reconstructing the complete kinematics of the reaction channels to
discriminate all background processes, such as multi-nucleon absorptions
and final state interactions.
An exclusive measurement is performed with the in-flight $^3$He$(K^-,n)$
reaction, which allows us to investigate the $K^-pp$ bound state
both in the formation via missing-mass spectroscopy and its decay via
invariant-mass spectroscopy using the emitted neutron and the
expected decay, $K^-pp \to \Lambda p \to \pi^-pp$, respectively.
The incident $K^-$ momentum of 1.0 GeV/$c$ is chosen to
maximize the $K^- N$ reaction rate~\cite{PDG}.

If both the binding energy and width of the $K^-pp$ bound state are assumed
to be approximately 100~MeV/$c^2$, as indicated by recent experimental
results~\cite{exp-Knucl}, the missing-mass resolution of the $(K^-,n)$
reaction is required to be less than 10~MeV/$c^2$ ($\sigma$) to
discriminate the $K^-pp$ signal from physical backgrounds such as the
two nucleon absorption processes from $K^- + ^3$He interactions.
For the invariant-mass spectroscopy, a large acceptance detector
surrounding the target system is essential.
To separate the expected $K^-pp$ decay modes of $\Lambda + p$ and
$\Sigma^0 + p$, the detector is designed to identify secondary charged
particles from the target and to reconstruct the $K^-pp \to \Lambda + p$
decay with an invariant-mass resolution of less than 20~MeV/$c^2$ ($\sigma$).
Additional information on the other decay channels such as $p + \Sigma +
\pi$ will also be studied.

\subsection{Precision spectroscopy of Kaonic helium-3 3d $\to$ 2p X-rays
  (the E17 experiment)}
The aim of the E17 experiment is to determine the shift and width of the
2p state of kaonic $^3$He and $^4$He atoms with a precision better than
1 eV~\cite{E17}.
In a recent theoretical calculation, the 2p level shift of kaonic $^3$He
and $^4$He atoms was calculated by a coupled channel scheme with a
phenomenological deep potential, which accommodates the
existence of deeply bound kaonic nuclear states~\cite{KatomA}.
In contrast to the calculations with optical models, which provide a very
small 2p level shift below 1 eV~\cite{Katom}, the coupled-channel
calculation claims the possibility of a large 2p level shift ($\sim$ 10
eV) for kaonic $^3$He and/or $^4$He atoms, depending on the strength of
the $\bar K$-nucleus potential.
Therefore, precision measurements of the energy shift of the 2p level of
kaonic $^3$He and $^4$He are a matter of great interest and can impose
the most stringent constraint on the $\bar K$-nucleus strong interaction
parameters.

By stopping negative kaons in a liquid helium target and using
high-resolution silicon drift X-ray detectors (SDDs), it is possible to
measure the strong-interaction shift of 3d $\to$ 2p X-rays from kaonic
$^3$He and $^4$He atoms. 
To carry out the measurement of the X rays with the world's
highest precision, i.e., better than~1 eV , we will achieve better energy
resolution and higher statistics than those of the E570 experiment at
KEK~\cite{E570} and the SIDDHARTA experiment at LNF~\cite{SIDDHARTA}.
In addition, we will apply the techniques that worked for E570: 
in-beam energy calibration with fluorescence X-rays and background
reduction by a reaction vertex reconstruction.

\subsection{Spectroscopic study of hyperon resonances below the $\bar{K}N$
  threshold via the $(K^-, n)$ reaction on the deuteron (the E31
  experiment)}
The primary goal of the E31 experiment is exclusively to show the
spectral function of the $\Lambda(1405)$ resonance produced in the $\bar
K N \to \pi \Sigma$ channel via the in-flight $(K^-,n)$ reaction on the
deuteron~\cite{E31}.
Despite many experimental attempts and theoretical analysis
related to the $\Lambda(1405)$, the most fundamental unsettled question
remains, ``is the $\Lambda(1405)$ located at 1405 MeV/$c^2$ or at 1420
MeV/$c^2$?''. 
So far, in the Review of Particle Physics by the Particle Data
Group~\cite{PDG2010}, the existence of $\Lambda(1405)$ was
established using only M-matrix analysis by R. H. Dalitz {\it et
al.}~\cite{Dalitz}, in which the estimated mass of the $\Lambda(1405)$ is
1406.5 $\pm$ 4.0 MeV/$c^2$ with a full width of 50 $\pm$ 2 MeV/$c^2$
based on the $K^-p \to (\pi^-\Sigma(1670)^+ \to \pi^-\pi^+\Lambda(1405))
\to \pi^-\pi^+\pi^-\Sigma^+ \to \pi^-\pi^+\pi^-\pi^+n$ reaction data
samples in the hydrogen bubble chamber at a kaon momentum of 4.2
GeV/$c$~\cite{exp-lam3}.
In addition to the above, the new review of 2012~\cite{PDG} adopted $M =
1405 ^{+1.4}_{-1.0}$ from J. Esmaili {\it et al.}~\cite{Esmaili} based on
the old bubble chamber data of stopped $K^-$ in $^4$He~\cite{exp-lam1}.

Theoretically, the resonance has been interpreted as an $I=0$
quasi-bound state, embedded in the $\Sigma-\pi$ continuum.
In recent years, strongly attractive $\bar{K} N$ interactions were
deduced from a coupled-channel approach based on the ansatz that the
$K^-$ bound state is located at 1405 MeV/$c^2$, and were used to
predict strongly bound and dense $\bar{K}$ nuclear states~\cite{AY}.
On the other hand, recent analyses based on the chiral unitary model
claim that $\Lambda(1405)$ has a two-pole structure;
one is the $\pi\Sigma$ state and the other is the $\bar KN$ state, which
results in a resonance position of about 1420 MeV/$c^2$ in the $\bar K N
\to \pi \Sigma$ channel and thus a much shallower binding scheme,
predicting much more weakly bound $\bar{K}$ systems~\cite{Lambda1405}.  
This controversy should be solved by the new experiments.

To clarify the nature of $\Lambda(1405)$, decomposition of
$\Lambda(1405)$ states coupled to $\bar KN$ is essential.
To this end, in E31, precise measurements of the $K^-d \to \pi\Sigma n$
reaction (where production is directly coupled to the $\bar K N$
interaction channel) will be used to study the nature of
$\Lambda(1405)$.
Missing-mass spectra in the inclusive $(K^-,n)$ reaction on the deuteron
target will be measured with a resolution of less than 10~MeV/$c^2$
($\sigma$).
A beam momentum of 1.0 GeV/$c$ will be used for the reasons outlined
in the E15 section.
All the  $\Sigma^+ \pi^-$, $\Sigma^- \pi^+$, and $\Sigma^0 \pi^0$ final
states must be identified.
Therefore, the experimental requirements of E15 also satisfy those of
E31.
In addition to the E15 requirement, E31 also requires the detection and
identification of a backward-going proton from the $\Lambda(1405) \to
\pi^0 \Sigma^0$ decay followed by $\Sigma^0 \to \gamma \Lambda \to \gamma
\pi^- p$, because the $\Lambda(1405)$ recoils at a backward angle,
i.e., the angular distribution of decay protons is boosted backward.

\section{Detector overview}
\begin{figure}[htbp]
   \begin{center}
    \includegraphics[width=\columnwidth]{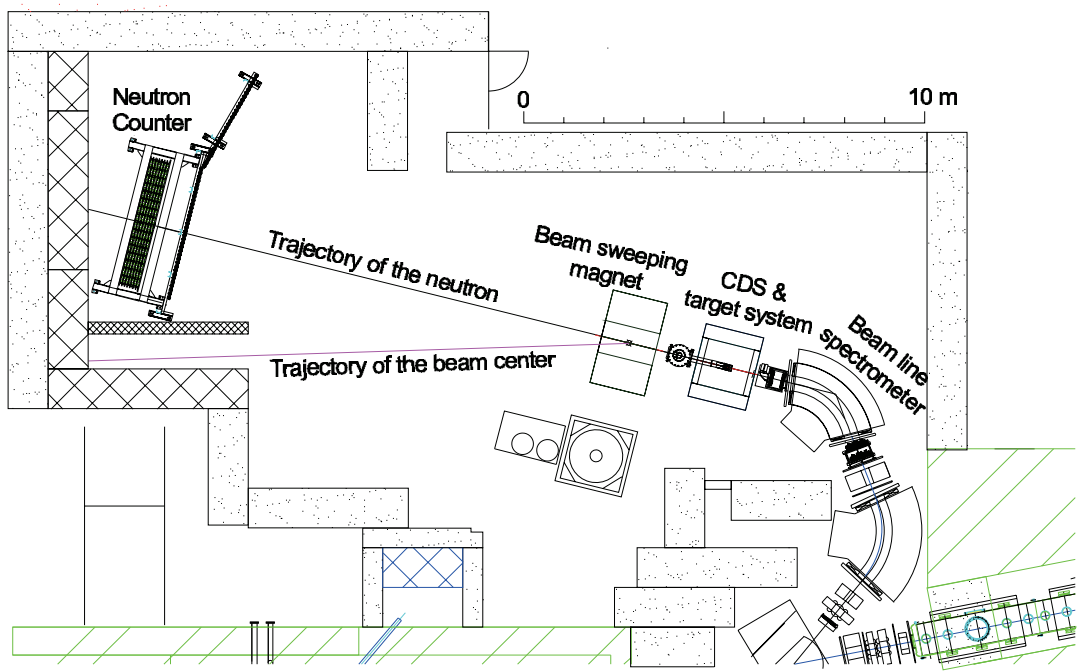}
    \caption{Schematic view of the K1.8BR spectrometer.
    The spectrometer consists of a beam line spectrometer, a
    cylindrical detector system that surrounds the liquid
    $^3$He/$^4$He/D$_2$ target system to detect the decay particles from
    the target region, a beam sweeping magnet, and a neutron
    time-of-flight counter located $\sim$15~m downstream from the target
    position.}
    \label{fig:spectrometer}
   \end{center}
  \end{figure}

A dedicated spectrometer was designed and constructed at the K1.8BR beam
line to satisfy all the experimental requirements described in the
previous section.
The spectrometer consists of a high precision beam line spectrometer, a
liquid $^3$He/$^4$He/D$_2$ target system, a cylindrical detector system
(CDS) that surrounds the target to detect the decay particles from the
target region, and a neutron time-of-flight counter array located
$\sim$15~m downstream from the target position, as shown in
Fig.~\ref{fig:spectrometer}.
In the following, a brief overview of the detector system at K1.8BR
is provided.

\begin{figure}[htbp]
   \begin{center}
    \includegraphics[width=0.8\columnwidth]{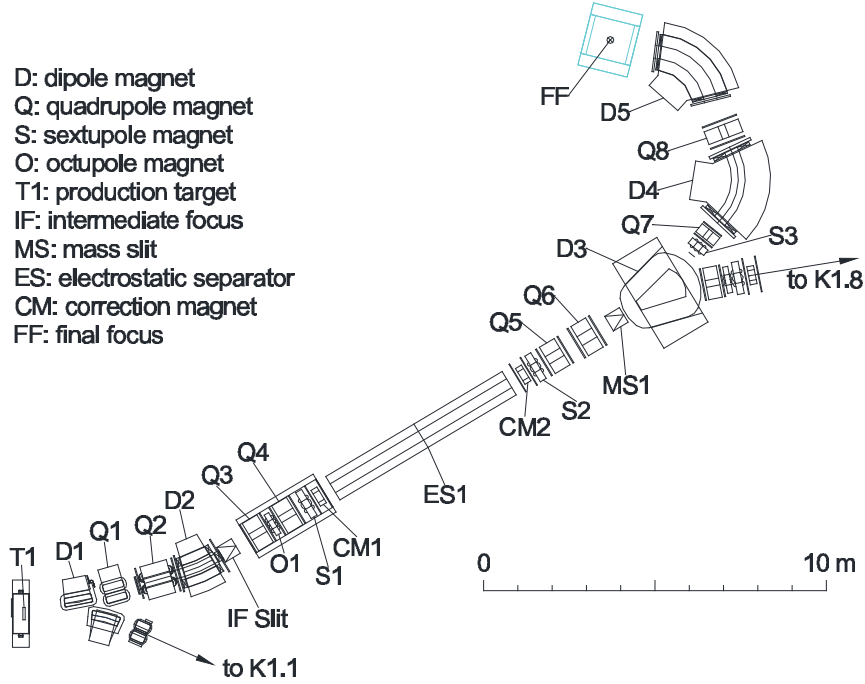}
    \caption{Configuration of the K1.8BR beam line in the hadron hall
    of J-PARC.}
    \label{fig:K18BR}
   \end{center}
  \end{figure}

The secondary beam line K1.8BR was constructed at the hadron hall
of the J-PARC 50~GeV proton synchrotron (PS).
The 30~GeV primary proton beam accelerated by the PS is transported to
the hadron hall through the beam switching yard and is focused on the
secondary-particle-production target, T1.
The K1.8BR beam line branches off from K1.8 at a bending magnet
downstream of an electrostatic separator, ES1, used to purify secondary
beams of charged particles with momenta up to 1.2 GeV/$c$ in the K1.8BR
beam line.
The configuration is shown in Fig.~\ref{fig:K18BR} and its parameters
are summarized in Table~\ref{tab:K18BR}.
The length of only 31.3~m optimizes the transport of low-momentum kaons.
The intensity of a 1.0~GeV/$c$ $K^-$ beam is expected to be $8 \times
10^4$ per second based on an estimation using the Sanford-Wang
formula~\cite{SanfordWang}.
An operational beam power of 270~kW (30~GeV, 9~$\mu$A proton beam)
striking a 54~mm thick nickel target (30\% loss target) was assumed.
Details of the beam line, such as the beam line elements and optical
design, are described in Ref.~\cite{Ieiri}.

\begin{table}[htbp]
  \begin{center}
   \caption{Parameters of the K1.8BR beam line.}
   \label{tab:K18BR}
   \begin{tabular}{ll}
    \hline \hline
    Primary beam & 30 GeV/$c$ proton \\
    Repetition cycle & 6 sec\footnotemark[1] \\
    Flat top & 2.93 sec\footnotemark[1]\\
    \hline
    Production target & Pt(50\% loss) / Ni(30\% loss) \\
    Production angle & 6 degrees \\
    Length (T1-FF) & 31.3 m \\
    Momentum range & 1.2 GeV/$c$ max. \\
    Acceptance & 2.0 msr$\cdot$\% ($\Delta\Omega\cdot\Delta p/p$)\\
    Momentum bite & $\pm$ 3\% \\
    \hline
   \end{tabular}
   \end{center}
\end{table}

The beam line spectrometer is composed of beam line magnets, trigger
counters, beam trackers, and a kaon identification counter.
The beam line magnets, composed of an SQDQD system, are located
downstream of a branching magnet D3.
The beam trigger is a coincidence signal from two trigger counters
located downstream from magnets D3 and D5, separated by a 7.7~m flight
path.\footnotetext[1]{Parameters in February 2012.}
Kaon beams ranging in momentum from 0.9 to 1.2~GeV/$c$ are identified
with the kaon identification counter.
Pions in the beams are discriminated from kaons using the kaon
identification counter, and protons are removed by ES1.
The trajectory of the kaon beam is tracked with the two beam line chambers
installed across the D5 magnet.
The kaon momentum is analyzed using this tracking information
together with beam optics of the D5 beam line magnet to attain an
expected momentum resolution of $\sim$0.1\%.
The beam line spectrometer was completed in January 2009, when the first
beam was delivered to the J-PARC hadron hall.

The liquid $^3$He/$^4$He target system, whose design is based on the
techniques developed for the $^4$He target used by KEK-PS E471, E549,
and E570, is located at the final focus.
In E17, the stopped-$K^-$ experiment, X-ray measurements are performed
using the SDDs installed in the target chamber.
For E31, a liquid D$_2$ target system similar to the $^3$He/$^4$He
system has been developed.
The liquid $^3$He/$^4$He target system was completed in 2008, and the
SDDs were ready in 2010.

Decay particles from the target are detected by the CDS, which
consists of a solenoid magnet, a cylindrical drift chamber (CDC),
and a cylindrical detector hodoscope (CDH).
The CDS has a solid angle coverage of 59\% of 4$\pi$.
Detailed tracking information on charged particles is obtained from the 
CDC, which operates in a solenoidal magnetic field of 0.7~T.
Particle identification is obtained using time-of-flight (TOF) together
with the trigger counter.
The basic CDS system was completed in 2008, and new detectors for
the CDS upgrade have been developed.

A forward neutron generated by the in-flight $(K^-, n)$ reaction is
detected by a forward neutron TOF counter array.
TOF distance, optimized to be $\sim$15~m from the target, will enable us
to achieve a total missing-mass resolution of less than 10 MeV/$c^2$
($\sigma$) with $\sim$150 ps TOF resolution of the system.
To perform efficient on-line particle identification of forward
neutral particles by the neutron TOF counter, the incident beam that
passes through the target is bent by a sweeping magnet placed just after
the CDS.
The neutron TOF counter and the sweeping magnet were installed in
2012.

\section{Beam line detectors}
\begin{figure}[htbp]
   \begin{center}
    \includegraphics[width=0.6\columnwidth]{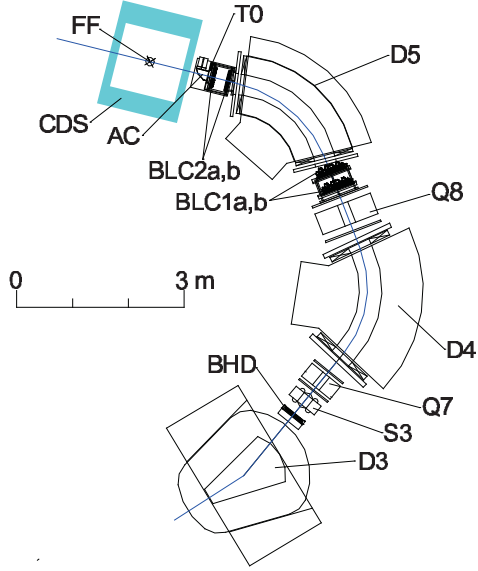}
    \caption{Schematic view of the beam line spectrometer, which
    consists of trigger counters (BHD and T0), beam line
    chambers (BLC1 and BLC2), and a kaon identification counter (AC).}
    \label{fig:beamline}
   \end{center}
  \end{figure}

A schematic view of the beam line spectrometer is presented in
Fig.~\ref{fig:beamline}.
It is composed of beam line magnets, trigger counters, beam trackers,
and a kaon identification counter.
The beam trigger is generated by a coincidence signal of a beam hodoscope
detector (BHD) and a time zero counter (T0); the flight length between
the BHD and T0 is 7.7~m.
The kaon beam with momentum around 1.0 GeV/$c$ is identified by using an
aerogel Cherenkov counter (AC) with a refractive index of 1.05.
The kaon beam is tracked with two beam trackers -- a beam line chamber 1
(BLC1) and a beam line chamber 2 (BLC2) -- and the momentum of the kaon
is analyzed with this tracking information together with the beam optics of
the D5 beam line magnet.
These basic detectors are common for all the experiments at the K1.8BR
beam line.

In addition to the above detectors, an energy measurement counter (E0)
and a vertex beam line drift chamber (VBDC) are used in the
stopped-$K^-$ experiment E17.
To stop the kaons in the target, a kaon beam of 0.9
GeV/$c$ is degraded by carbon and copper blocks placed after T0.
The degraded kaons are transferred into the target through E0 and the
VBDC, and stopped inside the target. 
The reaction vertex is obtained from an incident kaon and an outgoing
secondary charged particle track reconstructed by the VBDC and the CDC,
respectively.
By applying a correlation cut between the reaction vertex in the beam
direction and the energy loss in E0, in-flight kaon decay/reaction
events are rejected and continuum background events are drastically
reduced as a result.

\subsection{Trigger counters}
The BHD and T0 are segmented plastic scintillation counters
located downstream of the D3 and the D5 magnet, respectively.
The T0 signal is used as the event time-zero signal.

The BHD has an effective area of 400~mm (horizontal) $\times$ 160~mm
(vertical) segmented into 20 units horizontally, and T0 is 160~mm
(horizontal) $\times$ 160~mm (vertical) segmented into 5 units
horizontally.
To avoid over-concentration of the beam on one segment, T0 is
rotated by 45 degrees.
The BHD scintillator is made of Saint-Gobain BC412 with a unit size of
160~mm (height) $\times$ 20~mm (width) $\times$ 5~mm (thickness).
The unit size of the Saint-Gobain BC420 scintillator in T0 is 160 mm
(height) $\times$ 32~mm (width) $\times$ 10~mm (thickness).
In both of the trigger counters, the scintillation light is transferred
to a pair of 3/4~inch Hamamatsu H6612B photomultipliers that are
attached to the top and bottom ends.
Since the coincidence rate of the top and bottom photomultipliers will
reach $\sim$1~M counts per spill, the high voltage bleeders of all the
photomultipliers are modified to supply adequate current to the last three
dynodes.
Discriminated signals from the top and bottom photomultipliers
are coincidenced and provide the timing of each segment.
The typical TOF resolution between the BHD and T0 is 160~ps ($\sigma$)
after a slewing correction is applied.

\subsection{Kaon identification counter}
For the kaon beam trigger, the AC located downstream of T0 is used to
identify the kaon.
Kaons in the momentum region from 0.9 GeV/$c$ to 1.2 GeV/$c$, used for
the experiments at K1.8BR, are clearly separated from pions.

\begin{figure}[htbp]
   \begin{center}
    \includegraphics[width=0.6\columnwidth]{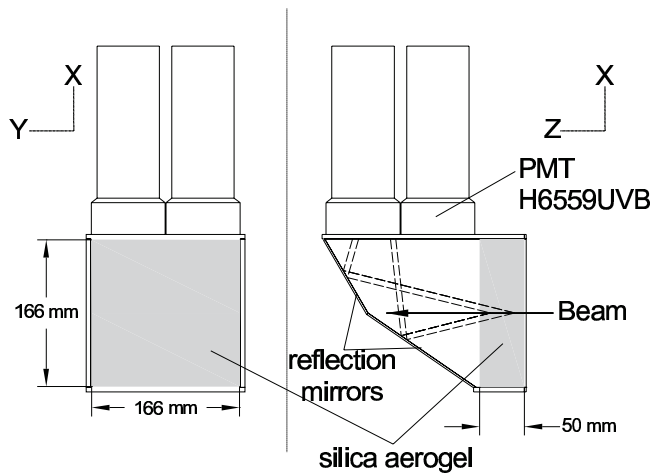}
    \caption{Schematic drawings of the aerogel Cherenkov counter.
    Typical beam trajectories and Cherenkov light paths are shown.}
    \label{fig:AC}
   \end{center}
  \end{figure}

The AC, which uses SP-50 silica aerogel produced by Matsushita Electric
Works, has an effective area of 166~mm (width) $\times$ 166~mm (height)
$\times$ 50~mm (thickness).
Cherenkov photons radiated in the beam direction are reflected by forward
optical mirrors and read out by four photomultipliers as shown in
Fig.~\ref{fig:AC}.
The single-photon-sensitive Hamamatsu H6559UVB photomultipliers have a
3~inch diameter photocathode on UV-transparent glass windows.

\begin{figure}[htbp]
 \begin{minipage}{0.5\textwidth}
  \begin{center}
   \includegraphics[width=\columnwidth]{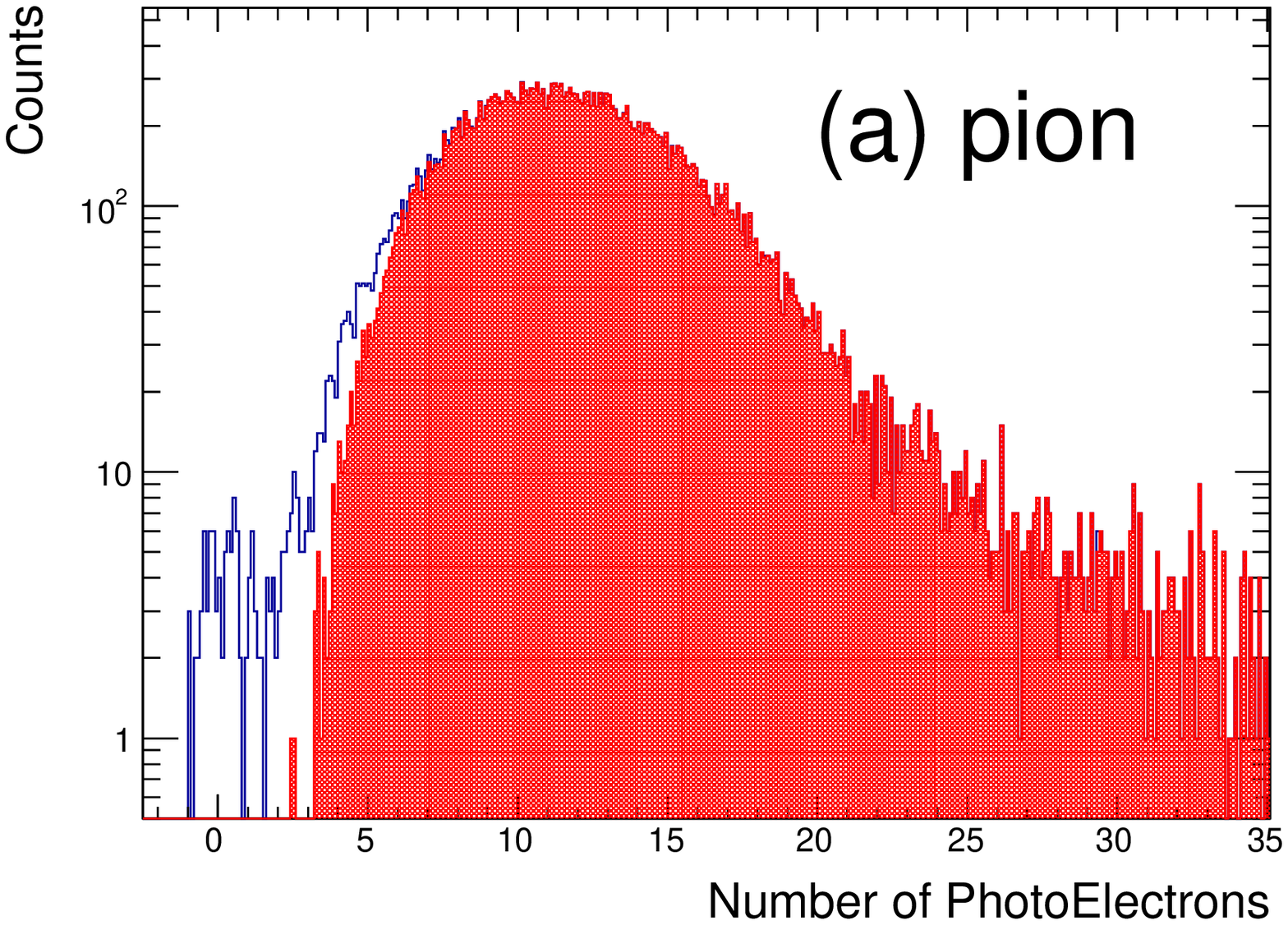}
  \end{center}
 \end{minipage}
 \begin{minipage}{0.5\textwidth}
  \begin{center}
   \includegraphics[width=\columnwidth]{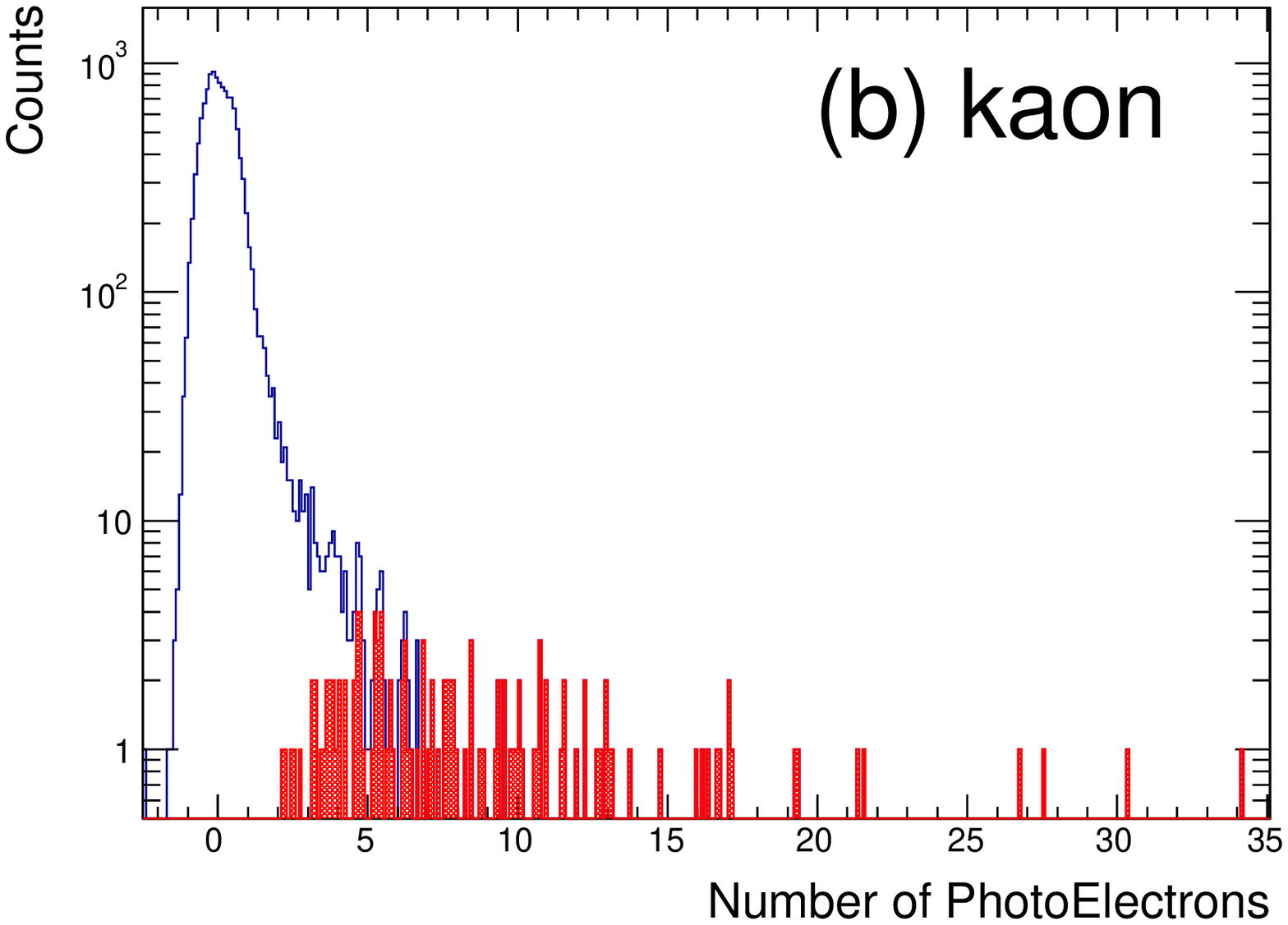}
  \end{center}
 \end{minipage}
 \caption{AC Pulse height distributions for 1.0 GeV/$c$ (a) pions and
 (b) kaons.
 Solid histograms represent (a) pions and (b) kaons identified
 with the TOF between the BHD and T0.
 Filled histograms represent the events identified as pions by the AC. 
 }
 \label{fig:AC_ADC}
\end{figure}

AC pulse height distributions for 1.0 GeV/$c$ pions and kaons, which
are obtained by summing ADC spectra of the four photomultipliers, are
shown in Fig.~\ref{fig:AC_ADC}.
In the figures, particle identifications are achieved by the TOF method
between the BHD and T0 in off-line analysis.
On-line pion identification is performed by the AC at a threshold level
of $\sim$5 photoelectrons.
A pion detection efficiency of 97\% is achieved, whereas the
missidentification ratio of a kaon as a pion is 1\%.

\subsection{Beam line chambers}
Kaon beam tracking is performed with two similar planar drift chambers,
BLC1 and BLC2, installed across the D5 magnet.

BLC1 consists of two sets of the same design of drift chamber, BLC1a and 
BLC1b, which have 8 layers with a $UU'VV'UU'VV'$ configuration.
In the $U$ and $V$ layers the wires are tilted by $\pm$ 45 degrees.
Each layer contains 32 sense wires with a drift length of 4~mm
corresponding to an effective area of 256~mm $\times$ 256~mm.
The number of readout channels is 256 for both BLC1a and BLC1b, which are
installed 300~mm apart upstream of the D5 magnet.

BLC2 is similar to BLC1; BLC2 consists of two sets of the same
drift chamber, BLC2a and BLC2b.
Each chamber has a $UU'VV'UU'VV'$ configuration and 32 sense wires per
layer, i.e, the number of readout channels is 256 for both BLC2a and
BLC2b.
In the $U$ and $V$ layers the wires are tilted by $\pm$ 45 degrees.
The drift length of 2.5~mm corresponds to an effective area of 160~mm
$\times$ 160~mm.
BLC2a and BLC2b are installed 275~mm apart downstream of the D5
magnet.

Both BLC1 and BLC2 use 12.5 $\mu$m diameter gold-plated tungsten wires
with 3\% rhenium and 75 $\mu$m diameter copper-beryllium wires for the
sense and potential wires, respectively.
The cathode planes are made of 12.5 $\mu$m aluminized Kapton.
The readout electronics of both chambers consist of a preamplifier
card with amplifier-shaper-discriminator ICs (ASD,
SONY-CXA3653Q~\cite{ASD}, $\tau$ = 16~ns) mounted on the chambers, an
LVDS-ECL converter, and a TDC.
The output signal of the ASD board is sent to the LVDS-ECL converter board
via 7~m long twisted-pair cables.
From the LVDS-ECL converter, the signal is transferred to the counting
house with 50~m long twisted-pair cables.
The chamber gas is an argon-isobutane mixture passed through a methylal
(dimethoxy-methane) bubbler at a refrigerator temperature of
4 $^\circ\mathrm{C}$ with a ratio of 76\% (Ar), 20\% (isobutane)
and 4\% (methylal).
The operating voltages of BLC1 and BLC2 are set at -1.25~kV on both the
potential wires and the cathode planes.
Typical position resolutions of 150~$\mu$m and detection efficiencies of
99\% for both BLC1 and BLC2 have been obtained.

\subsection{Detectors for the stopped-$K^-$ experiment}
E0, a segmented plastic scintillation counter that has an
effective area of 102~mm (horizontal) $\times$ 90~mm (vertical),
segmented into 3 units horizontally, is located just downstream of the
degraders.
The E0 scintillator is made of Eljen EJ-230 whose unit size
is 102~mm (height) $\times$ 30~mm (width) $\times$ 20~mm (thickness).
The scintillation light is transferred through light guides to a pair of
1~inch fine-mesh Hamamatsu H6152-01B photomultipliers attached to the
top and bottom ends.
The high voltage bleeders of the photomultipliers are modified to supply
additional current to the last three dynodes.

The VBDC planar drift chamber is located just before the target
system, hence was designed to be installable within the 30
cm inner diameter of the CDC.
It is 250~mm in diameter and 73.2~mm in height and consists of 8 layers
with a $UU'VV'UU'VV'$ configuration, where each layer contains 16 sense
wires with a drift length of 2.5~mm corresponding to an effective area
of 80~mm $\times$ 80~mm.
In the $U$ and $V$ layers the wires are tilted by $\pm$ 45 degrees.
The sense and potential wires are the same as those for the beam line
chambers.
The cathode planes are made of 7.5 $\mu$m thick Kapton, both
sides of which are coated with 0.1 $\mu$m aluminum with a layer of
0.0025 $\mu$m chromium as protection against oxidization.
The readout electronics and gas mixture are the same as those for the
beam line chambers.

For the in-flight experiments E15 and E31, E0 and VBDC are replaced
by the detectors for the CDS upgrade, the TGEM-TPC, the BPC, and the
BPD detectors, which are described later.

\section{Cryogenic target systems}
\input{intro.tex}
\subsection{Liquid $^3$He target system}
\input{lhe3.tex}

\subsection{Liquid D$_2$ target system}
\input{d2.tex}

\section{Silicon drift X-ray detector}
\input{sdd.tex}

\section{Cylindrical detector system}
\begin{figure}[htbp]
   \begin{center}
    \includegraphics[width=\columnwidth]{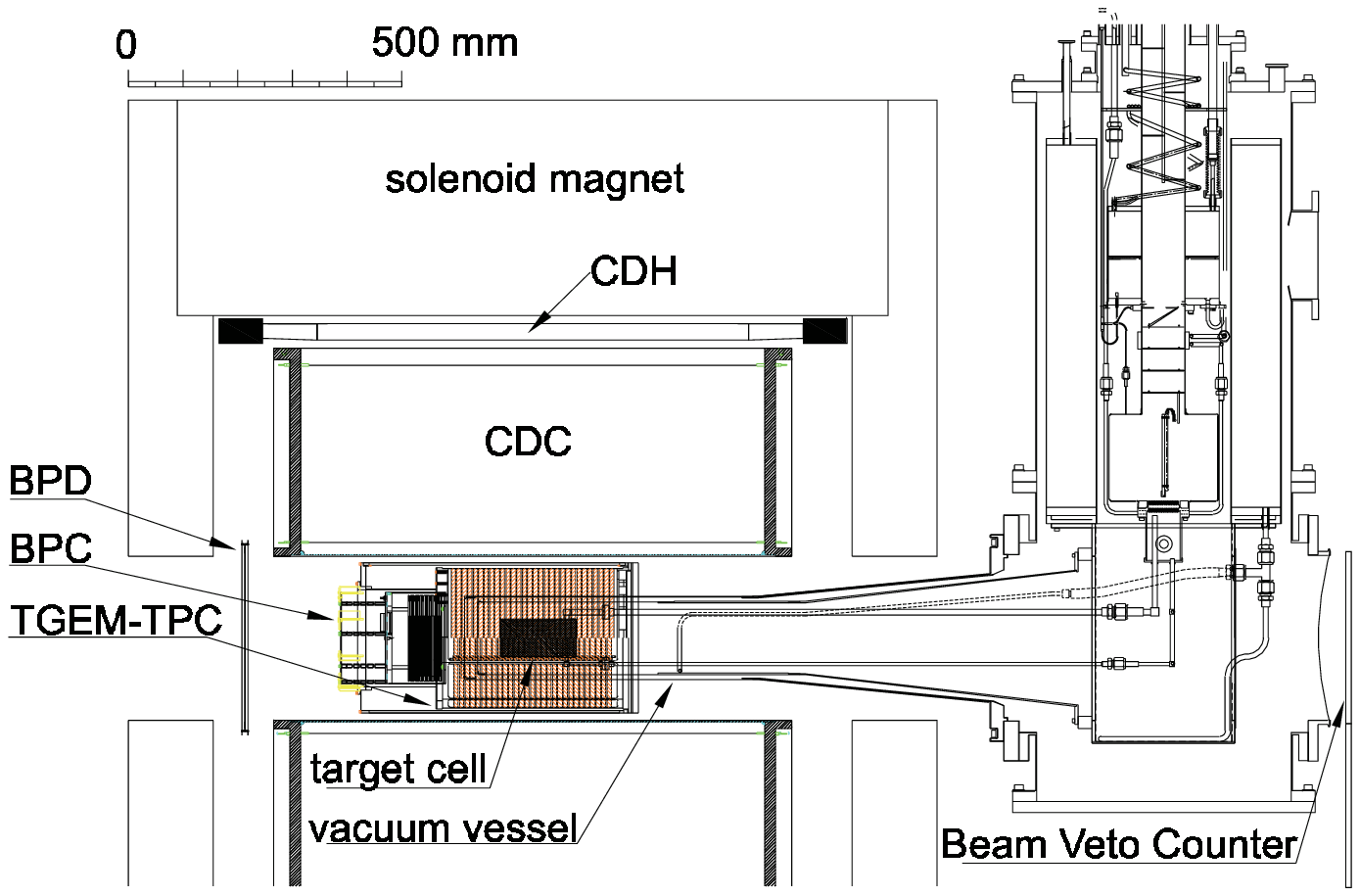}
    \caption{Schematic drawing of the CDS with the target system.}
    \label{fig:CDS}
   \end{center}
  \end{figure}

A schematic view of the cylindrical detector system (CDS) with the
target system is shown in Fig.~\ref{fig:CDS}.
The CDS consists of a solenoid magnet, a cylindrical drift
chamber (CDC), and a cylindrical detector hodoscope (CDH).
The decay particles from the target are reconstructed by the CDC, which
operates in a magnetic field of 0.7~T provided by the solenoid magnet.
The CDH is used for particle identification and as a charged particle
trigger.
The CDC and the CDH are common and basic detectors for all the
experiments at the K1.8BR beam line.

For in-flight experiments E15 and E31, additional detectors -- a time
projection chamber (TPC) with thick gas electron multipliers (TGEMs), a
backward proton chamber (BPC), and a backward proton detector (BPD) --
will be installed.
To realize an efficient measurement of the decay mode $K^-pp
\to p\pi\Sigma$, the TPC has been developed as an inner tracker for the
E15 upgrade that will be installed between the target and the CDC.
The BPC and the BPD have been developed aiming to reconstruct
backward-going particles that cannot be detected by the CDC.
The BPC and the BPD, which are installed just upstream of the target
system, enable measurement of a proton emitted from the $\Lambda(1405)
\to \pi^0 \Sigma^0$ decay mode ($\Sigma^0 \to \gamma \Lambda \to \gamma
\pi^- p$) by the time-of-flight method. 
The BPC and the BPD are also useful detectors for the E15 experiment,
because the requirement of a particle in the backward direction permits
an expansion of the acceptance in $\Lambda p n$ 3-body kinematical phase
space.

\subsection{Solenoid magnet}
\begin{figure}[htbp]
   \begin{center}
    \includegraphics[width=0.8\columnwidth]{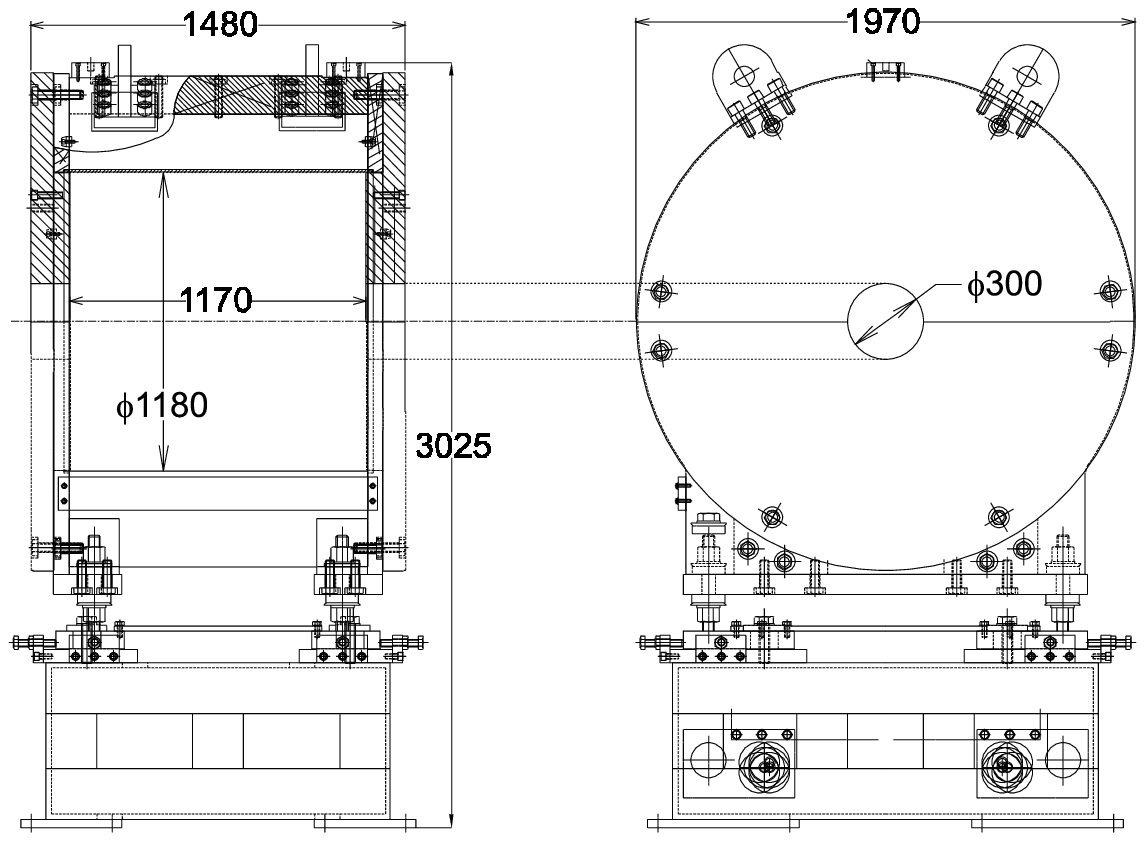}
    \caption{Design of the solenoid magnet (all dimensions in mm).}
    \label{fig:solenoid}
   \end{center}
  \end{figure}

The spectrometer magnet of the CDS is of a solenoidal type, whose bore
diameter is 1.18~m and whose length is 1.17~m with an overall
weight of 23 tons. 
The design of the solenoid magnet is shown in Fig.~\ref{fig:solenoid}.
It is located on the final focus point of the K1.8BR beam line.
The magnet provides a uniform field strength inside the tracking
volume ($|z| <$ 420~mm), whose strength is 0.7~T at the center of
the magnet.

\subsection{Cylindrical drift chamber}
\begin{figure}[htbp]
   \begin{center}
    \includegraphics[width=0.8\columnwidth]{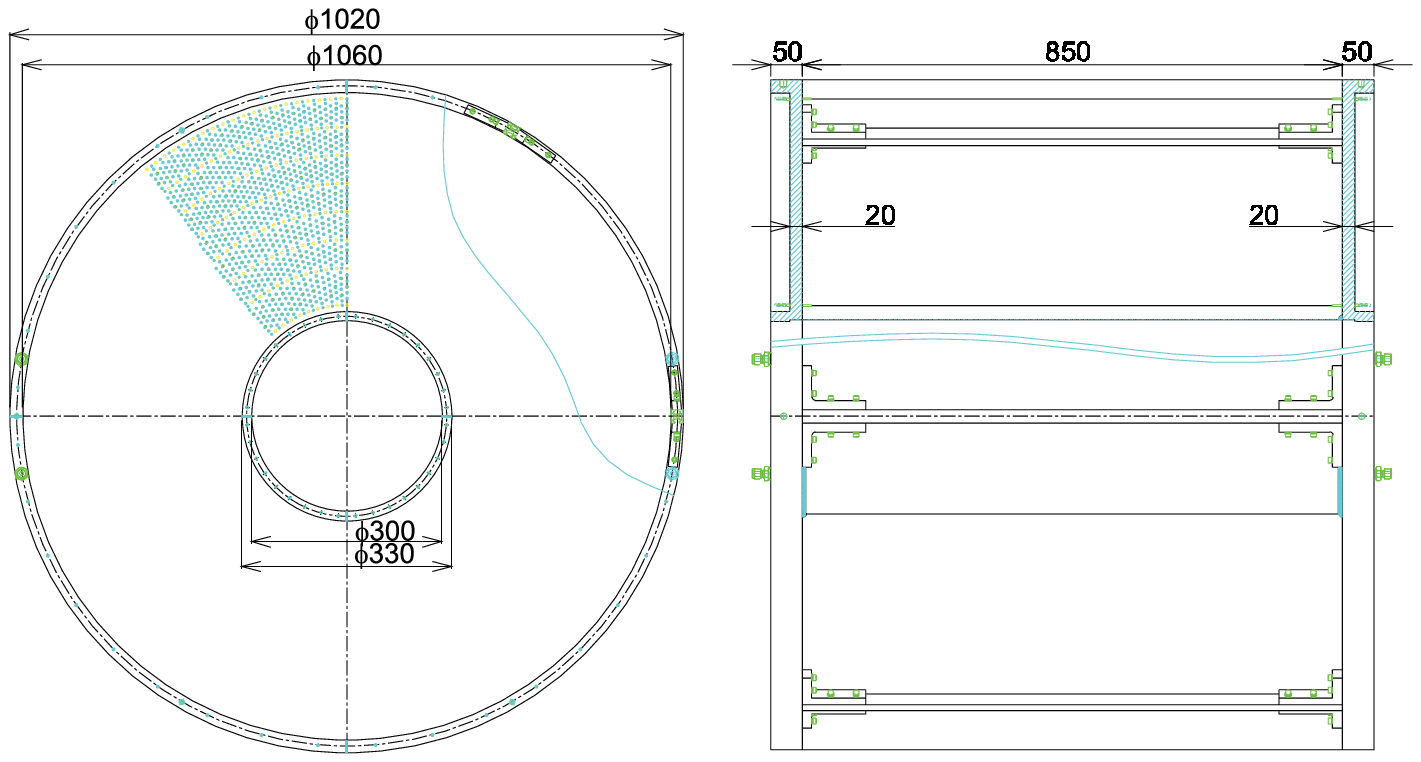}
    \caption{Design of the CDC (all dimensions in mm).
    The CDC consists of two aluminum end-plates, a 1~mm thick CFRP
    cylinder as an inner wall, and six aluminum posts that are placed
    outside the tracking volume.}
    \label{fig:CDC_structure}
   \end{center}
  \end{figure}

The CDC is a cylindrical wire drift chamber that contains 15 layers of
anode wires.
The structure of the CDC is shown in Fig.~\ref{fig:CDC_structure}.
The outer radius is 530~mm and the inner radius is 150~mm, with a total
length of 950~mm.
The wire length of axial layers is 838.8~mm, thus the angular coverage
is 49$^\circ$ $< \theta <$ 131$^\circ$ in the polar angle region
corresponding to a solid angle coverage of 66\% of 4$\pi$.

The CDC consists of two aluminum end-plates of 20~mm thickness, a
1~mm thick CFRP cylinder as the inner wall of the CDC, and six aluminum
posts that are placed outside the tracking volume.
The CDC uses gold-plated tungsten of 30~$\mu$m$~\phi$ for the sense
wires, and gold-plated aluminum of 100~$\mu$m$~\phi$ for the field and
guard wires.
These wires are supported by feedthroughs with a bushing inserted at
the end.
Bushes with an 80 and 200 $\mu$m$~\phi$ hole are used for the sense and
field/guard wires, respectively.

\begin{figure}[htbp]
   \begin{center}
    \includegraphics[width=0.5\columnwidth]{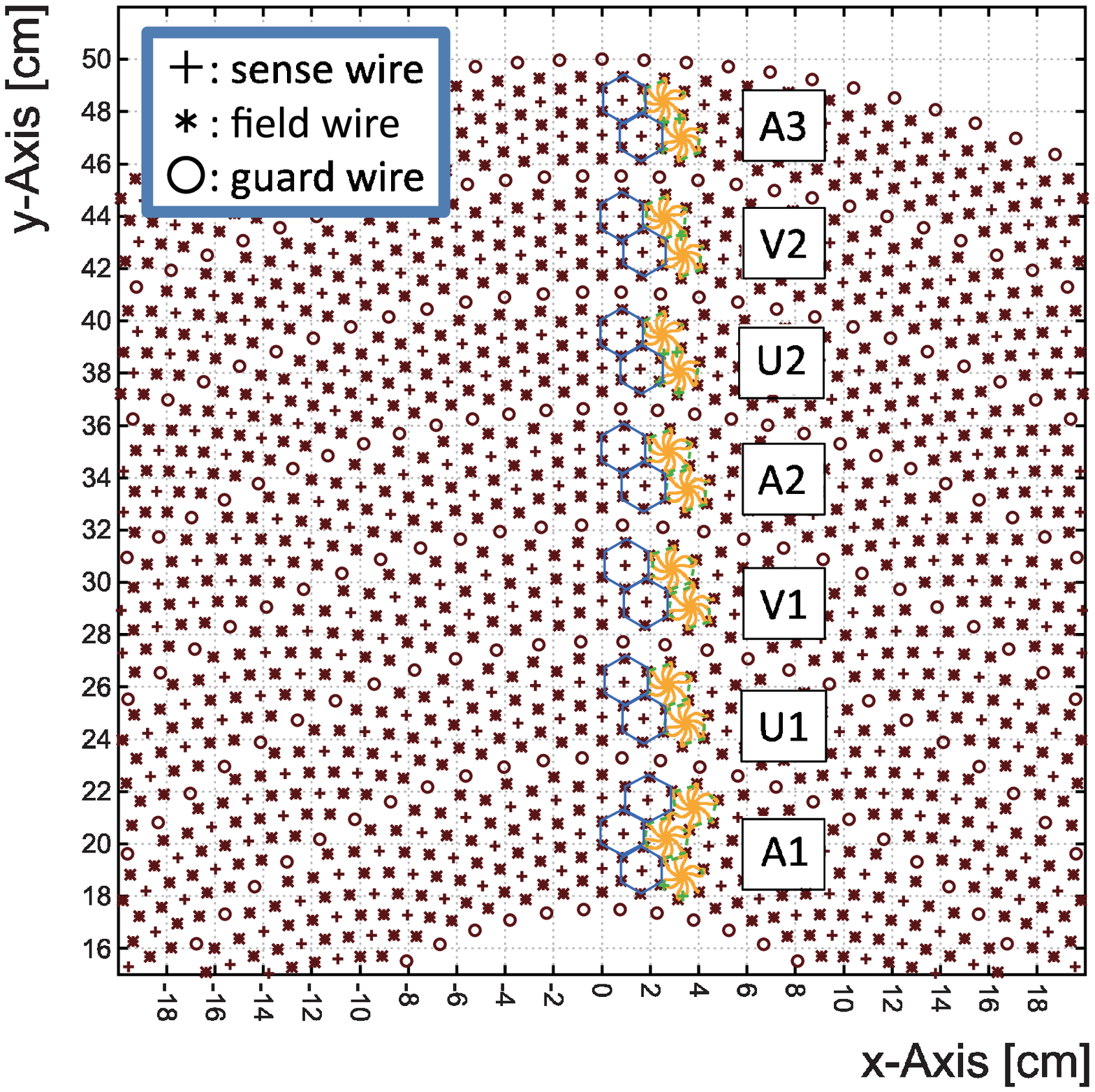}
    \caption{Cell structure of the CDC.}
    \label{fig:CDC_cell}
   \end{center}
  \end{figure}

The CDC has 15 layers of small hexagonal cells with a typical drift
length of 9~mm, which are grouped into 7 super-layers as shown in
Fig.~\ref{fig:CDC_cell}.
Table~\ref{tab:CDC} gives the detailed parameters of the wire
configuration.
The layers are in the radial region from 190.5~mm (layer \#1) to 484.5~mm
(layer \#15).
The 8 stereo layers tilted by about 3.5$^\circ$ are used to obtain
longitudinal position information.
The number of readout channels is 1816 and the total number of
wires in the CDC is 8064.

 \begin{table}[htbp]
  \begin{center}
   \caption{Wire configuration of the CDC.}
   \label{tab:CDC}
   \begin{tabular}{cccccccc}
    \hline \hline
    Super- & \multirow{2}{*}{layer} & Wire & Radius & Cell width & Cell
    width & Stereo angle & Signal channels\\
    layer &  & direction & (mm) & (degree) & (mm) & (degree) & per
    layer\\
    \hline
    \multirow{3}{*}{A1} & 1 & $X$ &  190.5 & \multirow{3}{*}{5.00} & 16.7 &
    0 & \multirow{3}{*}{72}\\
    & 2 & $X'$ & 204.0 & & 17.8 & 0 &\\
    & 3 & $X$  & 217.5 & & 19.0 & 0 &\\
    \hline
    \multirow{2}{*}{U1} & 4 & $U$ &  248.5 & \multirow{2}{*}{4.00} & 17.3 &
    -3.55 & \multirow{2}{*}{90}\\
    & 5 & $U'$ & 262.0 & & 18.3 & -3.74 &\\
    \hline
    \multirow{2}{*}{V1} & 6 & $V$ &  293.0 & \multirow{2}{*}{3.60} & 18.4 &
    3.77 & \multirow{2}{*}{100}\\
    & 7 & $V'$ & 306.5 & & 19.3 & 3.94 &\\
    \hline
    \multirow{2}{*}{A2} & 8 & $X$ &  337.5 & \multirow{2}{*}{3.00} & 17.7 &
    0 & \multirow{2}{*}{120}\\
    & 9 & $X'$ & 351.0 & & 18.4 & 0 &\\
    \hline
    \multirow{2}{*}{U2} & 10 & $U$ &  382.0 & \multirow{2}{*}{2.40} & 16.0 &
    -3.28 & \multirow{2}{*}{150}\\
    & 11 & $U'$ & 395.5 & & 16.6 & -3.39 &\\
    \hline
    \multirow{2}{*}{V2} &12 & $V$ &  426.5 & \multirow{2}{*}{2.25} & 16.7 &
    3.43 & \multirow{2}{*}{160}\\
    & 13 & $V'$ & 440.0 & & 17.3 & 3.54 &\\
    \hline
    \multirow{2}{*}{A3} & 14 & $X$ &  471.0 & \multirow{2}{*}{2.00} & 16.4 &
    0 & \multirow{2}{*}{180}\\
    & 15 & $X'$ & 484.5 & & 16.9 & 0 &\\
    \hline
   \end{tabular}
  \end{center}
 \end{table}

The drift gas is 1 atm of mixed argon (50\%)-ethane (50\%).
A high voltage is applied to the field and guard wires, and the sense
wires are kept at ground potential.
For the first super-layer (A1) and the second one (U1), a high voltage
of -2.8 kV is applied to the potential wires, and -2.7 kV to
the potential wires of the other super-layers.
In addition, -1.5 kV, -1.8 kV, and -0.6 kV are applied to the innermost, the
outermost, and the other guard wires, respectively.
The readout electronics of the CDC consists of a preamp card with ASDs
(SONY-CXA3653Q, $\tau$ = 16~ns), an LVDS-ECL converter, and a TDC --
the same as those for the beam line chambers. 

\begin{figure}[htbp]
 \begin{minipage}{0.6\textwidth}
  \begin{center}
   \includegraphics[width=\columnwidth]{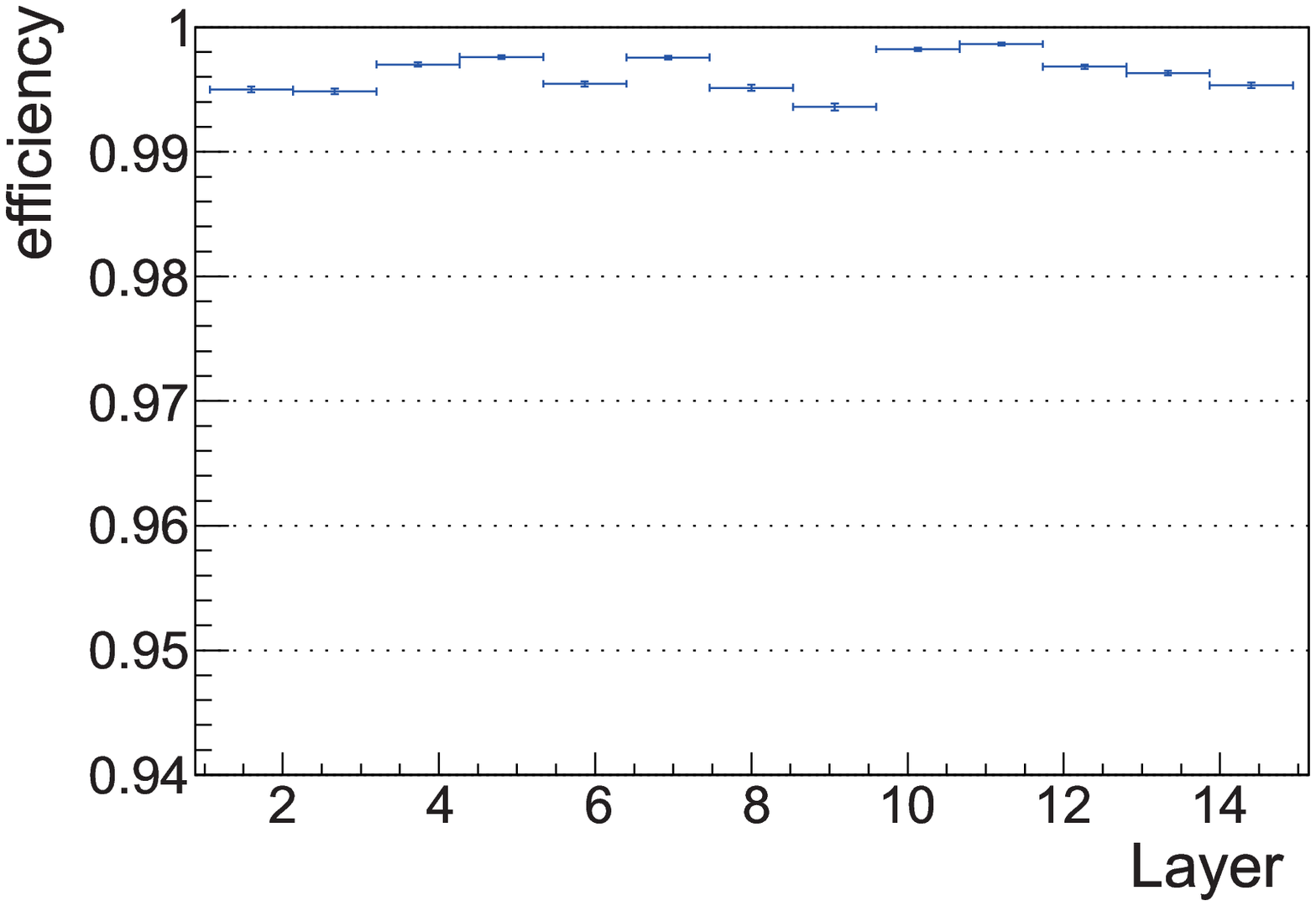}
   \caption{Layer efficiencies of the CDC measured in the engineering
   run conducted in June 2012.
   }
   \label{fig:CDCeff}
  \end{center}
 \end{minipage}
 \begin{minipage}{0.4\textwidth}
  \begin{center}
   \includegraphics[width=\columnwidth]{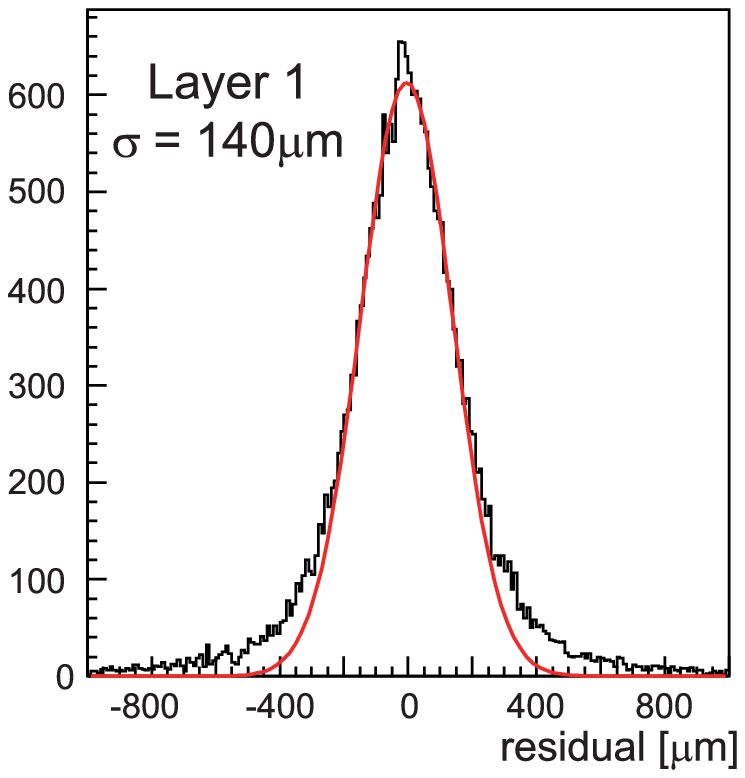}
   \caption{Typical residual distribution of the CDC.
   }
   \label{fig:CDCres}
  \end{center}
 \end{minipage}
\end{figure}

Figure~\ref{fig:CDCeff} shows the layer efficiencies of the CDC that were
obtained in the engineering run conducted in June 2012.
The efficiency is evaluated by (number of tracks having hits in the
relevant layer) / (number of tracks reconstructed except for hits in
the relevant layer).
All layers have efficiencies over 99\%.
In addition, the typical residual distribution of the CDC as obtained is shown
in Fig.~\ref{fig:CDCres}.
A typical intrinsic resolution of 200 $\mu$m was determined in the
data analysis.

\begin{figure}[htbp]
  \begin{center}
   \includegraphics[width=0.6\columnwidth]{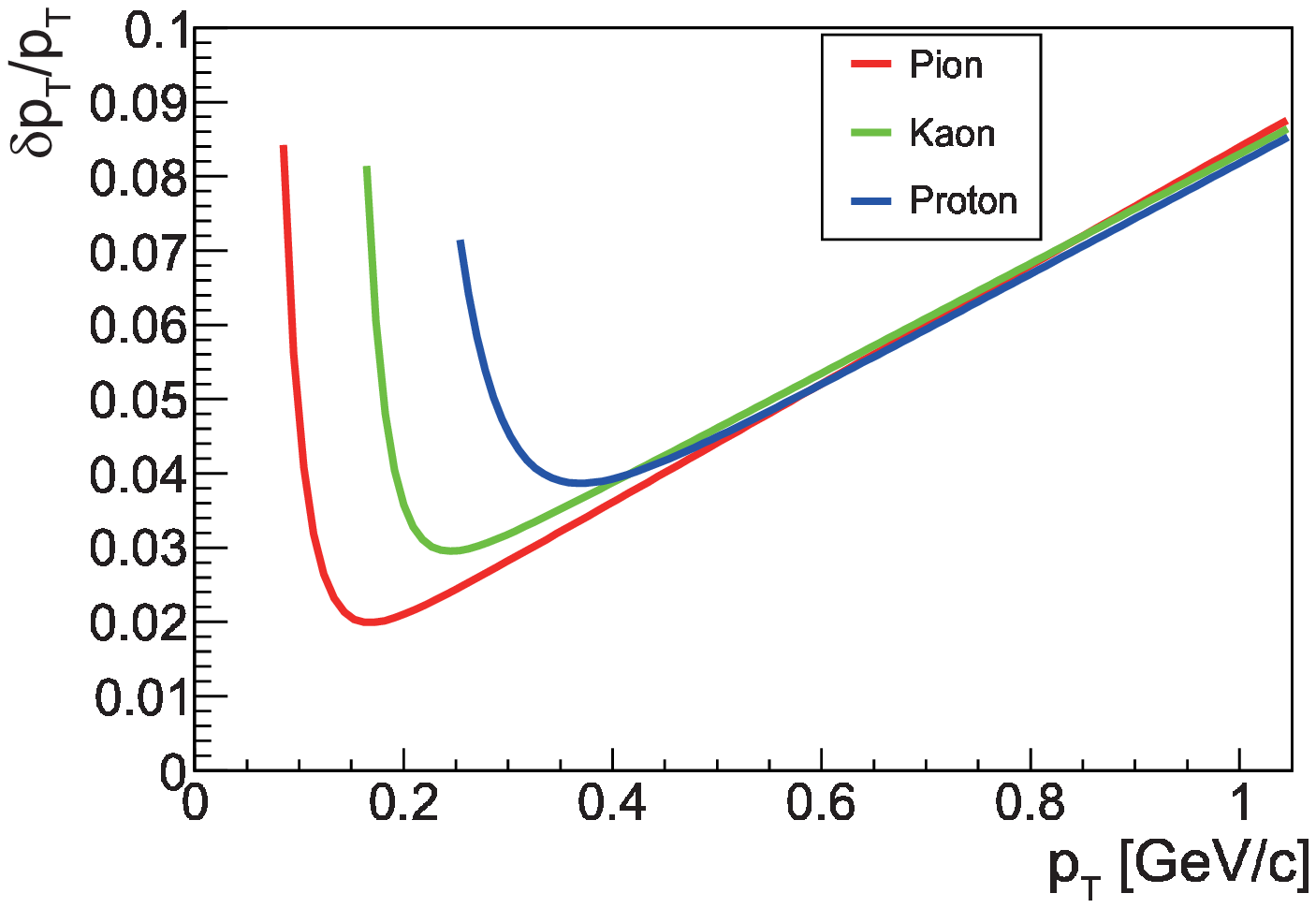}
   \caption{Momentum resolution of the CDC evaluated by the Monte Carlo
   simulation.
   The $p_t$ resolutions for pions, kaons, and protons are shown.
   }
   \label{fig:momres}
  \end{center}
\end{figure}

The momentum resolution of the CDC, shown in Fig.~\ref{fig:momres},
was estimated by a Monte Carlo simulation using the GEANT4
toolkit~\cite{GEANT4}.
In the simulation, the resolution of the CDC as well as energy loss and
scattering in detector materials were taken into account.
The calculated $p_t$ resolution was 8.4\% $p_t \oplus $ 1.1\%
$/ \beta$ and the estimated vertex resolution was 2.1~mm and 5.0~mm for the
perpendicular and parallel directions to the beam axis, respectively.

\subsection{Cylindrical detector hodoscope}
The CDH is a segmented plastic scintillation counter used for the charged
particle trigger and particle identification.
The CDH is located at a radius of 544~mm from the beam axis covering
a polar angle range from 54 to 126 degrees corresponding to a solid
angle coverage of 59\% of 4$\pi$.

The CDH consists of 36 modules, individually mounted on the inner wall
of the solenoid magnet.
The scintillators are made of Eljen EJ-200, with dimensions of 790~mm
in length, 99~mm in width, and 30~mm in thickness.
The scintillation light is transferred through light guides to a pair of
Hamamatsu R7761 fine-mesh 19-dynode photomultipliers 1.5 inches in
diameter.

The CDH is operated in the 0.7~T magnetic field with a typical PMT gain
of $\sim10^6$. 
The measured average time resolution of the CDH without a magnetic field
is 71 $\pm$ 3 ~ps ($\sigma$), obtained with cosmic ray data.

\subsection{TGEM time projection chamber}
The TPC with TGEMs has been developed to perform precise
vertex reconstruction for hyperon decays such as $\Lambda$ and
$\Sigma^{\pm}$, as an inner tracker of the CDS.
The TPC is installed between the target and the CDC.
For the TPC, the spatial resolution in the $Z$ direction should be less
than 1~mm, and the material budget in the CDC acceptance should
be minimized. 

  \begin{figure}[htbp]
   \begin{center}
    \includegraphics[width=\columnwidth]{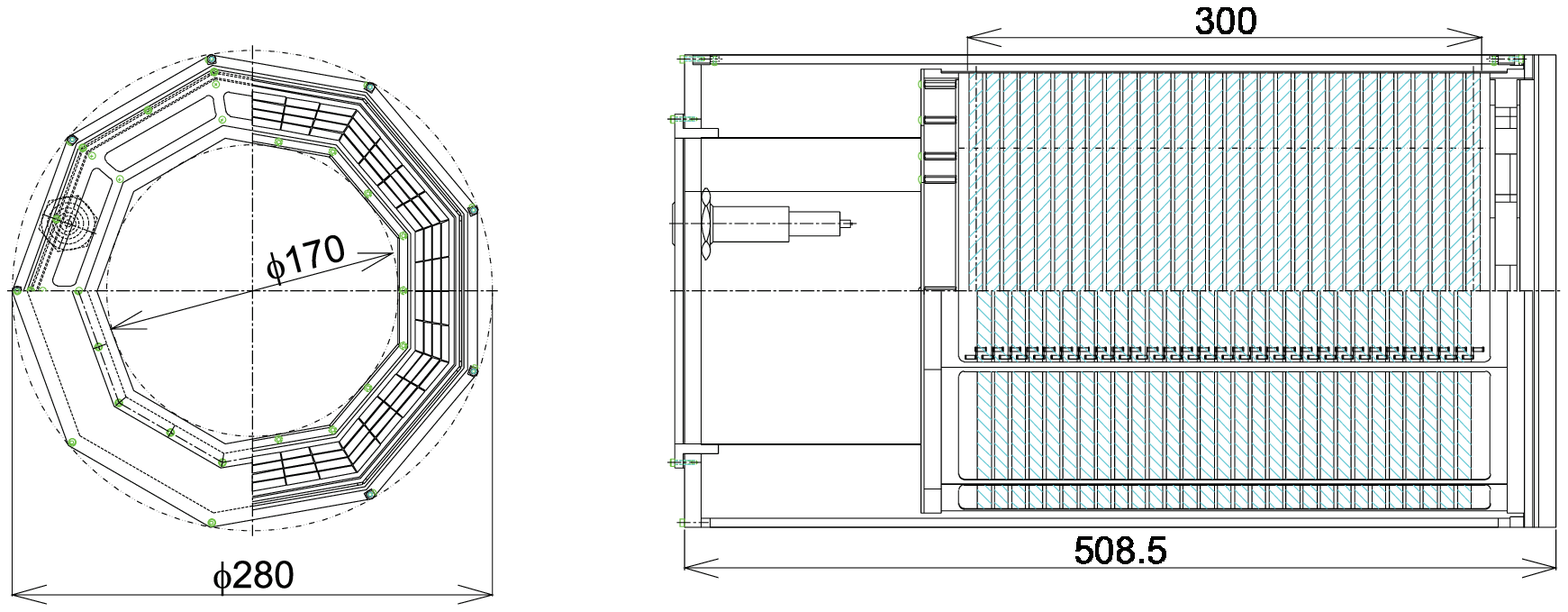}
    \caption{Design of the TGEM-TPC (all dimensions in mm).}
    \label{fig:TPC}
   \end{center}
  \end{figure}  

\begin{figure}[htbp]
   \begin{center}
    \includegraphics[width=0.6\columnwidth]{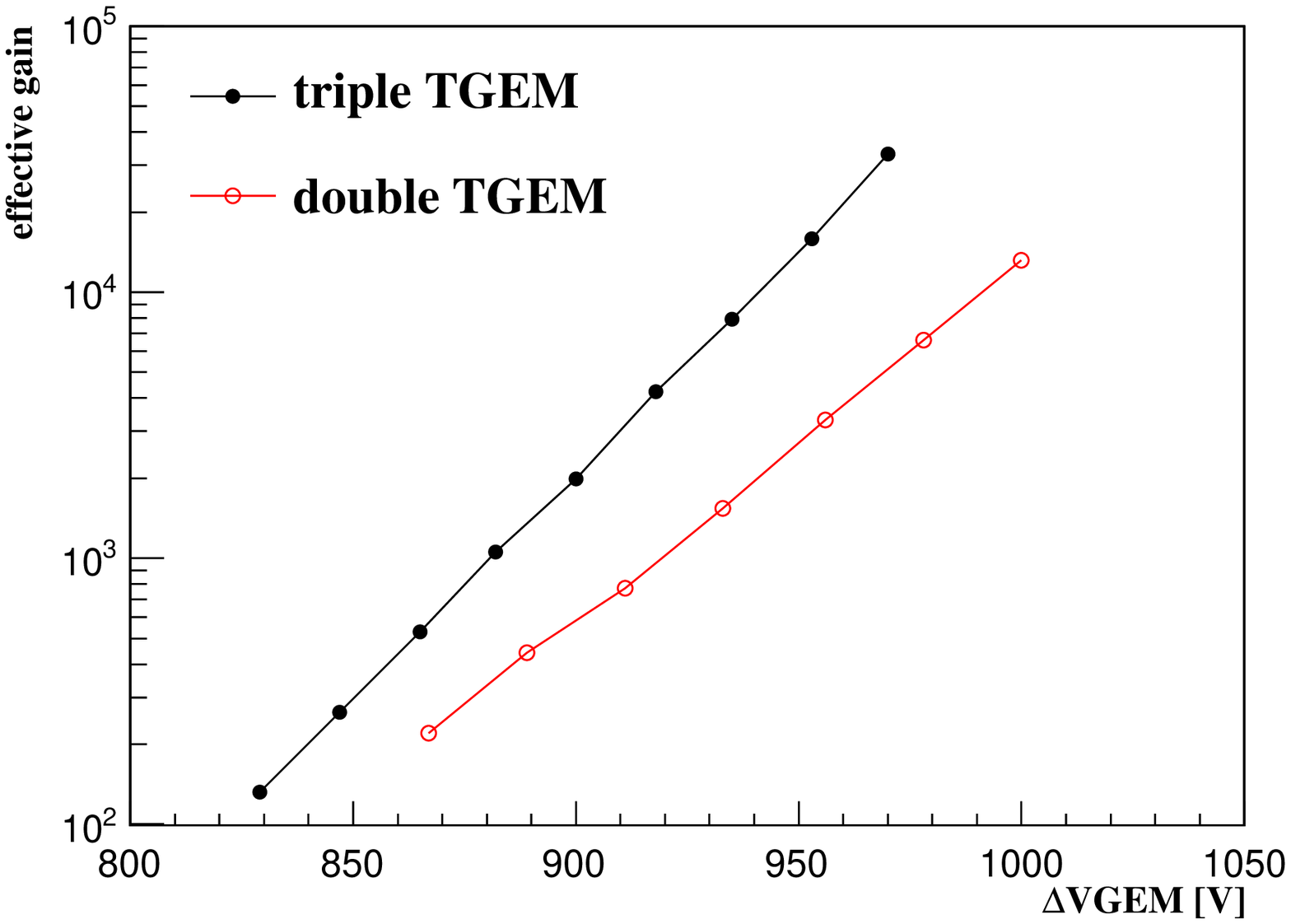}
    \caption{Effective gains of the nonagonal TGEMs with the graphite
    electrode as a function of the voltage across each TGEM.
    The data for the double- and triple-TGEM configurations are shown.
    }
    \label{fig:TPC-gain}
   \end{center}
  \end{figure}

The design of the TPC is shown in Fig.~\ref{fig:TPC}.
It is cylindrical in shape with inner diameter 170~mm, outer
280~mm, and is filled with P-10 gas (90\% Ar+ 10\% CH$_4$) at
atmospheric pressure.
The drift length is 30~cm in a field cage made from double-sided
flexible printed circuits (FPC) having staggered strip electrodes with a
width of 8~mm and pitch of 10~mm, connected with 1 M$\Omega$ resistors.
A drift field of 150 V/cm is applied to the field cages.
For amplification, a double or triple-TGEM structure is used; the TGEM
is economically constructed from double-clad 400 $\mu$m thick FR4 plate
using standard printed circuit board (PCB) techniques, and has
mechanically drilled holes~\cite{TGEM}.
To reduce both the energy and the propagation probability of
a discharge, the TGEM for the TPC has a nonagonal shape whose sides
are subdivided into three sectors and separately connected externally to
the voltage supply through high-value resistors.
High voltages are applied to the double/triple-TGEM resistor chain
through connectors penetrating the end cap.
The gain of these nonagonal TGEMs is greater than $10^4$ with graphite
electrodes having a resistance of about 30 $\Omega$/$\Box$ both with
double- and triple-TGEM configurations, as shown in
Fig~\ref{fig:TPC-gain}.
The graphite-electrode TGEM is a new development to protect the detector
and the readout electronics from damage by any occasional
discharge~\cite{RETGEM}.
The layout of the readout system is also nonagonal and is divided into 4
$\times$ 4 pads on each side with 4 mm long and 20 mm wide pads printed
on a standard PCB (the total number of readout channels is 144). 
For the TPC front-end electronics, preamplifier cards with ASDs (
SONY-CXA3653Q, $\tau$ = 80~ns) are used, i.e., the TPC provides only
beam-direction information on the tracks.

To evaluate the TPC performance, a test experiment using a
positron beam of around 600 MeV/$c$ was performed at the laser-electron
photon facility at SPring-8 (LEPS)~\cite{LEPS} in November 2011.
In the test experiment, the TPC was operated without the magnetic field
with only part of the readout system installed.
The typical spatial resolution obtained in the $Z$ direction was 
0.95 mm and 1.9 mm at drift length of 30 mm and 270 mm,
respectively.
A tracking efficiency of over 95\% was achieved independent of the drift
length.
The drift length dependence of the spatial resolution can be expressed as
$\sigma_{\rm z}^2 = \sigma_0^2+C_{\rm d}^2 \cdot z / N_{\rm eff}$, where
$\sigma_{\rm z}, \sigma_0, C_{\rm d}, z$, and $N_{\rm eff}$ are total
resolution, resolution
without diffusion, diffusion constant, drift distance, and effective
number of electrons, respectively~\cite{TPC}.
With a value of $C_{\rm d}$ = 0.34 mm and electric field of 150 V/cm in the
P-10 gas in an evaluation by the gas simulation program
Magboltz~\cite{Magboltz}, values of $\sigma_0$ = 0.73 mm and
$N_{\rm eff}$ = 7.7 were obtained.
When operating in a magnetic field, the resolution and efficiency are
expected to improve as a result of an increase in the effective
number of electrons.
This increase should achieve the design criteria.
The TPC will be completed in 2012 followed by a performance test in
which all of these basic parameters will be experimentally determined.

\subsection{Detectors for backward protons}
The BPC and BPD are installed just upstream of the target system
aiming to reconstruct backward-going particles, such as pions and
protons, which cannot be detected by the CDC.

\begin{figure}[htbp]
   \begin{center}
    \includegraphics[width=0.8\columnwidth]{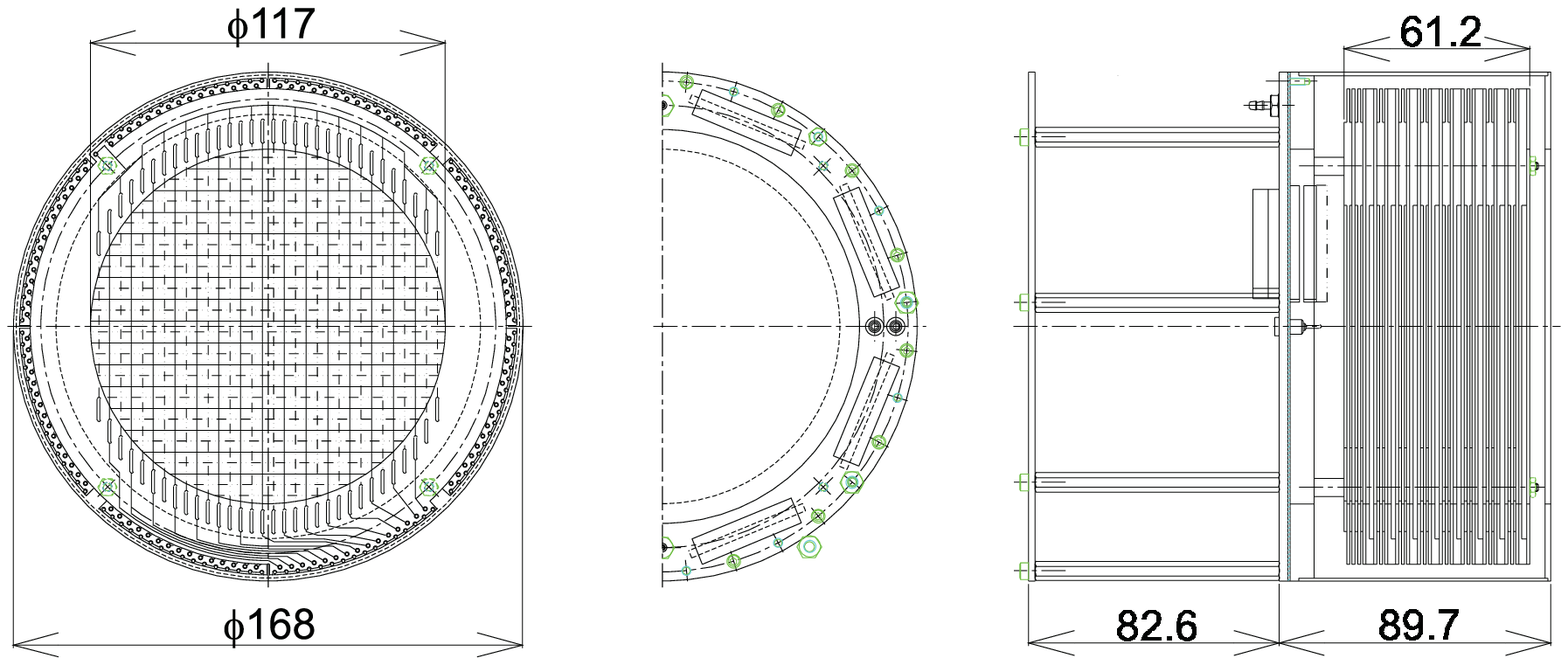}
    \caption{Design of the BPC (all dimensions in mm).}
    \label{fig:BPC}
   \end{center}
  \end{figure}

The BPC is a compact circular planar drift chamber located just
before the target system 168~mm in diameter and 89.7~mm in height.
Figure~\ref{fig:BPC} shows the design of the BPC, which consists of 8
layers with an $XX'YY'XX'YY'$ configuration, where the wires of the $Y$ layer
are tilted by 90 degrees.
Each layer contains 15 sense wires with a drift length of 3.6~mm
corresponding to an effective area with a 111.6~mm diameter.
The number of readout channels is 120.
The cathode planes are made of 9~$\mu$m carbon aramid foil, and the
sense and potential wires, readout electronics, and gas mixture of the
BPC are the same as those for the beam line chambers.
The operational voltage of the BPC is set at -1.45~kV on both the
potential wires and the cathode planes.
In the engineering run of 2012, individual layer efficiencies were
greater than 99\% and the position resolution was 150~$\mu$m.

The BPD is a plastic scintillator hodoscope array that is placed 0.5 m
upstream from the final focus, inside the solenoid magnet.
The size of the BPD is 350 mm (horizontal) $\times$ 340 mm (vertical)
segmented into 70 units.
Each scintillation counter, made of Eljen EJ-230, is 5~mm $\times$ 5~mm
$\times$ 340~mm.
Due to the strong magnetic field and a limited space, multi-pixel photon
counters (MPPC) with a 3 mm $\times$ 3 mm sensitive area were used
(Hamamatsu S10362-33-050C).
The MPPCs were put on both sides of each slab.
Signals from the MPPCs are read out by fast timing amps (ORTEC FTA820).
A typical time resolution of 160 ps was achieved in the engineering
run, which is sufficient to identify a particle in the backward
direction.

\section{Neutron time-of-flight counter and beam sweeping magnet}
The neutron TOF counter, placed $\sim$15~m away from the center of the
target at 0 degrees with respect to the beam direction, detects a
forward neutron generated by the in-flight $(K^-, n)$ reaction.
The neutron TOF counter is an array of scintillator counters previously
used by KEK-PS E549, which was reassembled for the E15 experiment.
The kaon beam is swept out from the acceptance of the neutron counter by
a sweeping magnet placed just after the CDS to perform
efficient on-line particle identification of forward neutral particles.
Charged particles within the acceptance of the neutron counter, such as
those generated on the wall of the beam sweeping magnet, are vetoed by a
charge veto counter located just upstream of the neutron counter.
In addition, a beam veto counter is installed between the CDS and the
beam sweeping magnet to reduce fake triggers caused by the decay of beam
kaons after the target.

\begin{figure}[htbp]
   \begin{center}
    \includegraphics[width=0.9\columnwidth]{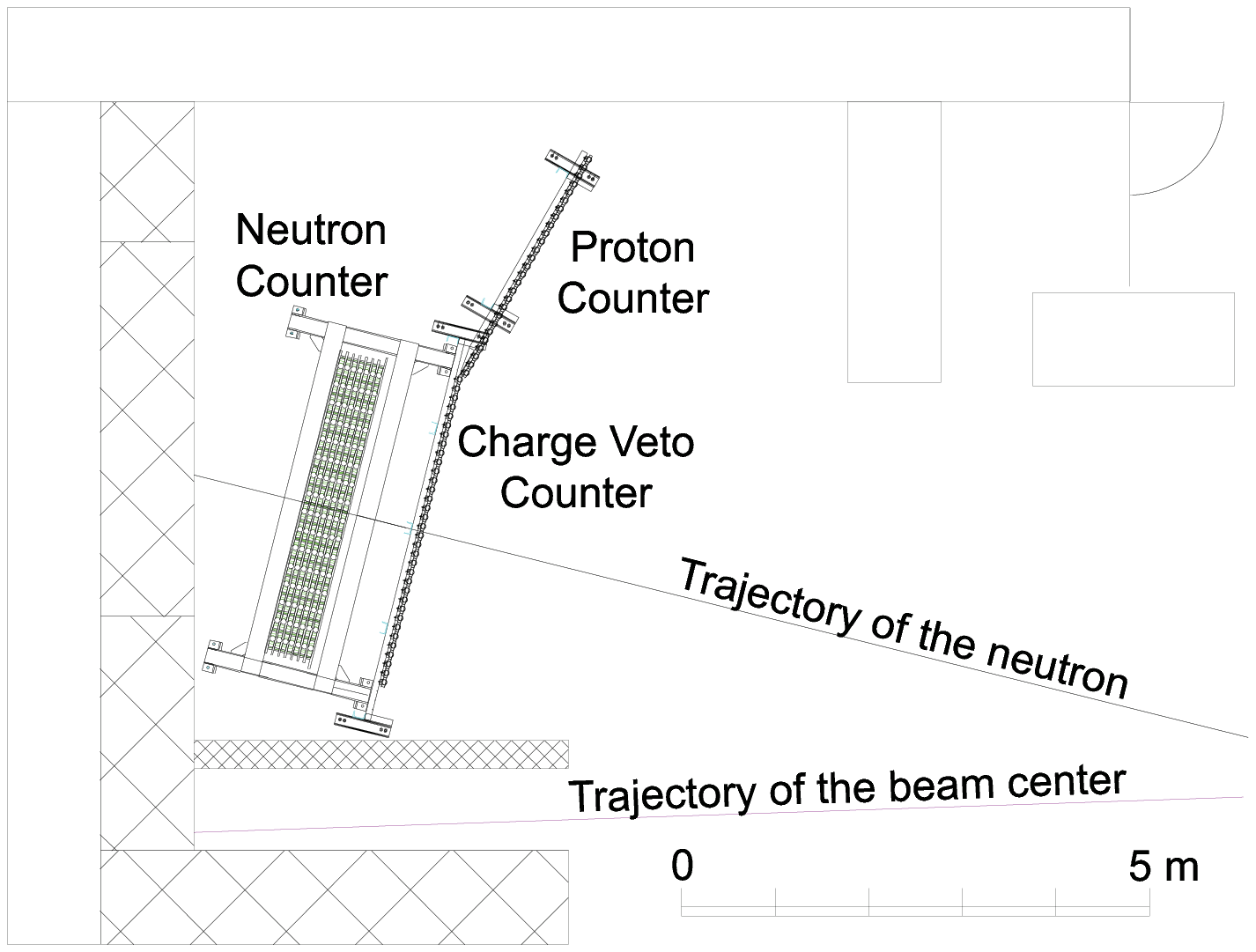}
    \caption{Schematic view of the neutron counter, the charge veto
    counter, and the proton counter.
    The neutron counter is located 14.7~m away from the final focus
    position.}
    \label{fig:NC}
   \end{center}
  \end{figure}

To measure both the $^3$He($K^-$, $p$) and the ($K^-$,
$n$) reactions, a proton TOF counter was installed.
A comparison of the two kinds of missing-mass spectra will provide
unique information on the isospin dependence of the kaon-nucleus ($\bar
K-NN$) interaction.
The proton counter is located alongside the charge veto counter, as an
extended wall of the charge veto counter, on the opposite side of a beam
dump to which the beam is bent by the beam sweeping magnet.
Figure~\ref{fig:NC} shows a schematic view of the neutron counter, the
charge veto counter, and the proton counter.

\subsection{Neutron time-of-flight counter}
A forward neutron generated by the in-flight $(K^-, n)$ reaction is
detected by the neutron TOF counter located 14.7~m away
from the final focus point in which the experimental target is installed.
The neutron TOF counter consists of an array of scintillation counters
and has an effective volume of 3.2~m (horizontal)
$\times$ 1.5~m (vertical) $\times$ 0.35~m (depth) segmented into
16-column (horizontal) $\times$ 7-layer (depth) units.
The acceptance of the neutron counter is $\pm$ 6.2$^\circ$ in the
horizontal direction and $\pm$ 2.9$^\circ$ in the vertical.
Each scintillation counter has dimensions of 20~cm (width) $\times$
150~cm (height) $\times$ 5~cm (thickness) viewed by 2~inch Hamamatsu H6410
photomultipliers attached to both long sides of the scintillator through
a Lucite light guide.
The scintillators for the first three layers are made of Saint-Gobain
BC408, and the other four layers are made of Saint-Gobain BC412.
The average time resolution of the neutron counter, measured with cosmic
rays, is 92 $\pm$ 10~ps ($\sigma$).
The detection efficiency for a $\sim$1 GeV/$c$ neutron is estimated to
be $\sim$ 35\% from the Monte Carlo simulation by the GEANT4 toolkit.

\subsection{Beam sweeping magnet}
A dipole magnet called Ushiwaka, which was used in the $\pi$2 beam line of
the 12 GeV proton synchrotron at KEK, is used as the beam sweeping
magnet.
It is located just downstream of the CDS.
The magnet has an aperture of 82~cm (horizontal) $\times$ 40~cm
(vertical) and a pole length of 70~cm, and is capable of providing a
maximum field of 1.6 T.
To sweep the beam away from the neutron counter acceptance
window and transfer the beam to the beam dump, the magnet provides a
field strength of $\sim$1.2 T at the center of the magnet.

\subsection{Charge veto counter}
The charge veto counter is located upstream of the neutron counter,
14.0~m away from the final focus point.
This counter is used as a charge veto counter for the neutron detector.
It has an effective area of 3.4~m (horizontal)
$\times$ 1.5~m (vertical) segmented into 34 units.
Each scintillation counter has dimensions of 10~cm (width) $\times$
150~cm (height) $\times$ 3~cm (thickness), and is equipped with two
2~inch Hamamatsu H6410 photomultipliers attached to
both long sides of the scintillator through a Lucite light guide.
The scintillators are of Eljen EJ-200 type.
The average time resolution measured with cosmic rays is 78 $\pm$ 7~ps
($\sigma$).

\subsection{Proton time-of-flight counter}
The proton TOF counter is installed as the extended wall of the
charge veto counter.
It has an effective area of 2.7~m (horizontal) $\times$ 1.5~m (vertical)
segmented into 27 units.
Each scintillation counter consists of a Saint-Gobain BC408 scintillator and
two Hamamatsu H6410 photomultipliers attached to both long sides of the
scintillator through a Lucite light guide. 
The average time resolution of the proton counter, obtained from cosmic ray
data, is 75 $\pm$ 6~ps ($\sigma$).

\subsection{Beam veto counter}
The beam veto counter is installed between the CDS and the beam
sweeping magnet.
The size of the beam veto counter is 315~mm (height) $\times$ 315~mm (width)
$\times$ 10~mm (thickness) made of Saint-Gobain BC408.
The scintillation light transferred through a light guide is read by a
2~inch fine-mesh Hamamatsu H6154 photomultiplier.
The high voltage bleeder of the photomultiplier is modified to supply
additional current to the last three dynodes.

\section{Trigger and data acquisition}

\subsection{Trigger}
To meet experimental requirements, a dedicated hardware trigger is applied
to each experiment.
The kaon beam trigger is common for all experiments, whereas the 
main triggers for the in-flight experiments (E15 and E31) and for
the stopped-$K^-$ experiment with the SDDs (E17) are substantially
different.
The E15 and E31 experiments use a 1.0 GeV/$c$ kaon beam, and detect
a forward neutron generated by the $^3$He or $d(K^-, n)$ reaction.
The E17 experiment uses a kaon beam of 0.9 GeV/$c$ that is stopped in
the liquid target by using degraders located just upstream of
the target system.
\begin{description}
 \item[(1) kaon beam trigger]\mbox{}\\
	    The elementary beam trigger is constructed by coincidence
	    signals from the beam line counters, the BHD and T0.
	    The kaon beam trigger $(K_{\rm beam})$ is selected from the beam
	    trigger by using the kaon identification counter, i.e., a
	    veto signal of the AC ($\overline{\rm AC}$) defines the kaon beam.
	    It is to be noted that (anti-) protons in the beam are
	    eliminated upstream of the beam line by using the ES1,
	    CM1, and CM2.
	    A logical expression of the kaon beam trigger is given as 
	    \begin{eqnarray} 
	     \nonumber
	      (K_{\rm beam}) \equiv ({\rm BHD}) \otimes ({\rm T0})
	      \otimes (\overline{\rm AC}).
	    \end{eqnarray}
 \item[(2) E15 main trigger]\mbox{}\\
	    A two-level trigger logic for the in-flight $^3$He$(K^-,
	    n$) reaction is applied.
	    To reconstruct the expected decay $K^-pp \to
	    \Lambda p \to p\pi^-p$ using the CDS, an event with two or
	    more CDH hits (${\rm CDH}_2$) is selected from the kaon beam
	    trigger in the first level (${\rm E15}_{\rm 1st}$).
	    In addition, no hit on the beam veto counter
	    ($\overline{\rm BVC}$) is required to reduce the trigger rate.
	    In the second level (${\rm E15}_{\rm 2nd}$), an event with a
	    forward neutron is chosen by requiring a neutron counter
	    hit ($\rm NC$) and a veto signal of the charge veto counter
	    ($\overline{\rm TOF}$).
	    The E15 main trigger is given as
	    \begin{eqnarray} 
	     \nonumber
	      ({\rm E15}_{\rm 1st}) &\equiv& (K_{\rm beam}) \otimes
	      ({\rm CDH}_2) \otimes
	      (\overline{\rm BVC}), \\
	     \nonumber
	      ({\rm E15}_{\rm 2nd}) &\equiv& ({\rm E15}_{\rm 1st})
	      \otimes ({\rm NC}) \otimes
	      (\overline{\rm TOF}).
	    \end{eqnarray}
 \item[(3) E17 main trigger]\mbox{}\\
	    A two-level trigger logic is applied to measure
	    X-rays from the kaonic helium-3 and -4 atoms.
	    The stopped-$K^-$ trigger ($K_{\rm stopped}$) is generated
	    by the signal from E0 and the veto-signal from
	    the beam veto counter with the kaon beam definition:
	    \begin{eqnarray} 
	     \nonumber
	      (K_{\rm stopped}) \equiv (K_{\rm beam}) \otimes ({\rm E0})
	      \otimes (\overline{\rm BVC}).
	    \end{eqnarray}
	    For the first-level trigger (${\rm E17}_{\rm 1st}$), one or
	    more CDH hits (${\rm CDH}_1$) is required to reduce the trigger
	    rate and obtain the reaction vertex.
	    A hit in the SDD is required in the second level, because
	    the timing of the SDD signal from the shaping amplifier is
	    too late for the trigger timing of the first level.
	    Therefore the E17 main trigger is
	    \begin{eqnarray} 
	     \nonumber
	      ({\rm E17}_{\rm 1st}) &\equiv& (K_{\rm stopped}) \otimes
	      ({\rm CDH}_1), \\
	     \nonumber
	      ({\rm E17}_{\rm 2nd}) &\equiv& ({\rm E17}_{\rm 1st})
	      \otimes ({\rm SDD}).
	    \end{eqnarray}
 \item[(4) E31 main trigger]\mbox{}\\
	    A two-level trigger logic for the in-flight $d(K^-,
	    n$) reaction is applied.
	    Since E31 measures $\Lambda(1405)$ decays into $\pi^0
	    \Sigma^0 \to \pi^0 \gamma \Lambda \to \pi^0 \gamma \pi^- p$
	    and $\pi^\pm \Sigma^\mp \to \pi^\pm \pi^\mp n$ using the CDS
	    in addition to the forward neutron, the trigger logic is the
	    same as that for the E15 trigger, except for a requirement for one
	    or more CDH hits in the first level.
	    Thus the E31 main trigger is
	    \begin{eqnarray} 
	     \nonumber
	      ({\rm E31}_{\rm 1st}) &\equiv& (K_{\rm beam}) \otimes
	      ({\rm CDH}_1) \otimes (\overline{\rm BVC}), \\
	     \nonumber
	      ({\rm E31}_{\rm 2nd}) &\equiv& ({\rm E31}_{\rm 1st})
	      \otimes ({\rm NC}) \otimes (\overline{\rm TOF}).
	    \end{eqnarray}
\end{description}

\subsection{Data acquisition system}
The on-line data acquisition system (DAQ) for the experiments at
K1.8BR consists of the TKO\cite{TKO}, VME, and PC Linux.
The signals from the detectors are fed into ADC and TDC modules slotted
into the TKO crates.
In the K1.8BR experiments, 10 TKO crates are used; they are parallel
read from the VME-SMP (super memory partner)\cite{SMP} via a TKO SCH (super
controller head).
The data stored in a buffer memory of the SMP is transferred to the
DAQ-PC through SBS Bit3 VME-to-PCI bridges.
Additionally, the E17 experiment uses another system that consists of
the VME (flash-ADC modules) and PC Linux for the readout of the SDDs.
The data is written to the disk on the DAQ-PC, and transferred to a PC
cluster server at RIKEN via the Internet.

\section{Spectrometer performance}
\begin{figure}[htbp]
  \begin{center}
   \includegraphics[width=0.5\columnwidth]{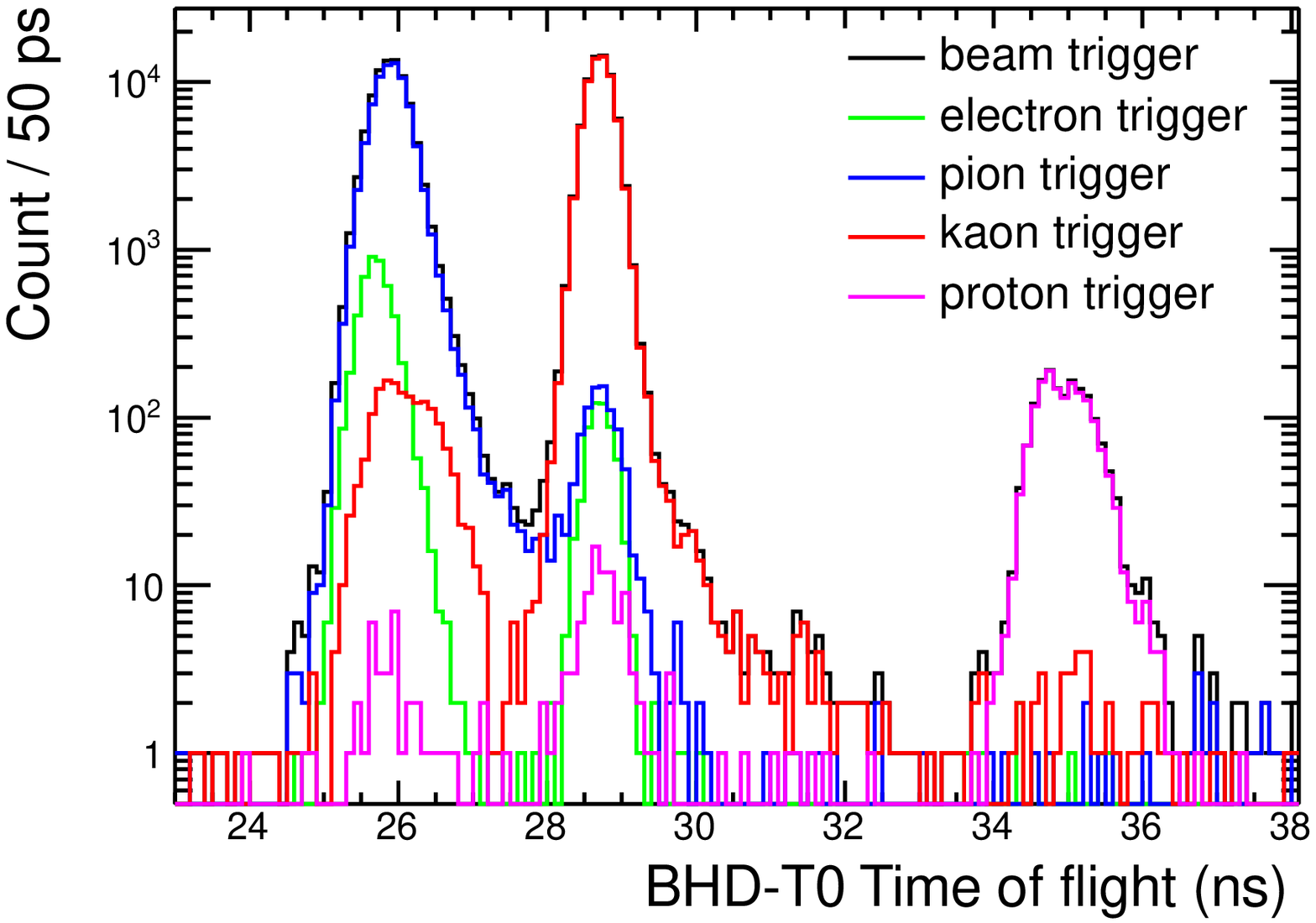}
   \caption{Time-of-flight spectrum between the BHD and T0 for 1.0
   GeV/$c$ positive particles with ES1 field of 50.0 kV/cm.
   Particle identification was performed using three Cherenkov
   counters.
   }
   \label{fig:beam-tof}
  \end{center}
\end{figure}

The quality of the secondary kaon beam is key for all of the experiments
at K1.8BR.
We have optimized the beam line since 2009 to obtain an intense and good
$K/\pi$-separated kaon beam by tuning the beam line spectrometer.
During a commissioning run in February 2012, beam line commissioning for
1.0~GeV/$c$ was accomplished and optimized parameters for the
spectrometer were obtained.
Details of the beam line performance are described in Ref.~\cite{Ieiri}.
It should be noted that the typical 1.0 GeV/$c$ kaon yield normalized by
an accelerator power of 1.0 kW was obtained to be 10 k / spill with a
$K/\pi$ ratio of 0.3, when a 60~mm thick platinum target (50\% loss
target) was used as the secondary-particle-production target, T1.
Figure~\ref{fig:beam-tof} demonstrates particle separation by
time-of-flight between the BHD and T0.
To perform on-line particle identification in the
commissioning run, two Cherenkov counters located downstream of the D5
magnet were used in addition to the AC; a gas Cherenkov counter (GC,
refractive index $n$ = 1.002) and a water Cherenkov counter (WC, $n$ = 1.33)
\footnote{
The on-line particle identification of electrons, pions, kaons, and
protons is performed with coincidence signals from $({\rm GC})$,
$(\overline{\rm GC}) \otimes ({\rm AC})$, $(\overline{\rm GC}) \otimes
(\overline{\rm AC}) \otimes ({\rm WC})$, and $(\overline{\rm GC})
\otimes (\overline{\rm AC}) \otimes (\overline{\rm WC})$,
respectively.}.
Each particle species is clearly separated in the time-of-flight
spectrum.

\begin{figure}[htbp]
  \begin{center}
   \includegraphics[width=0.5\columnwidth]{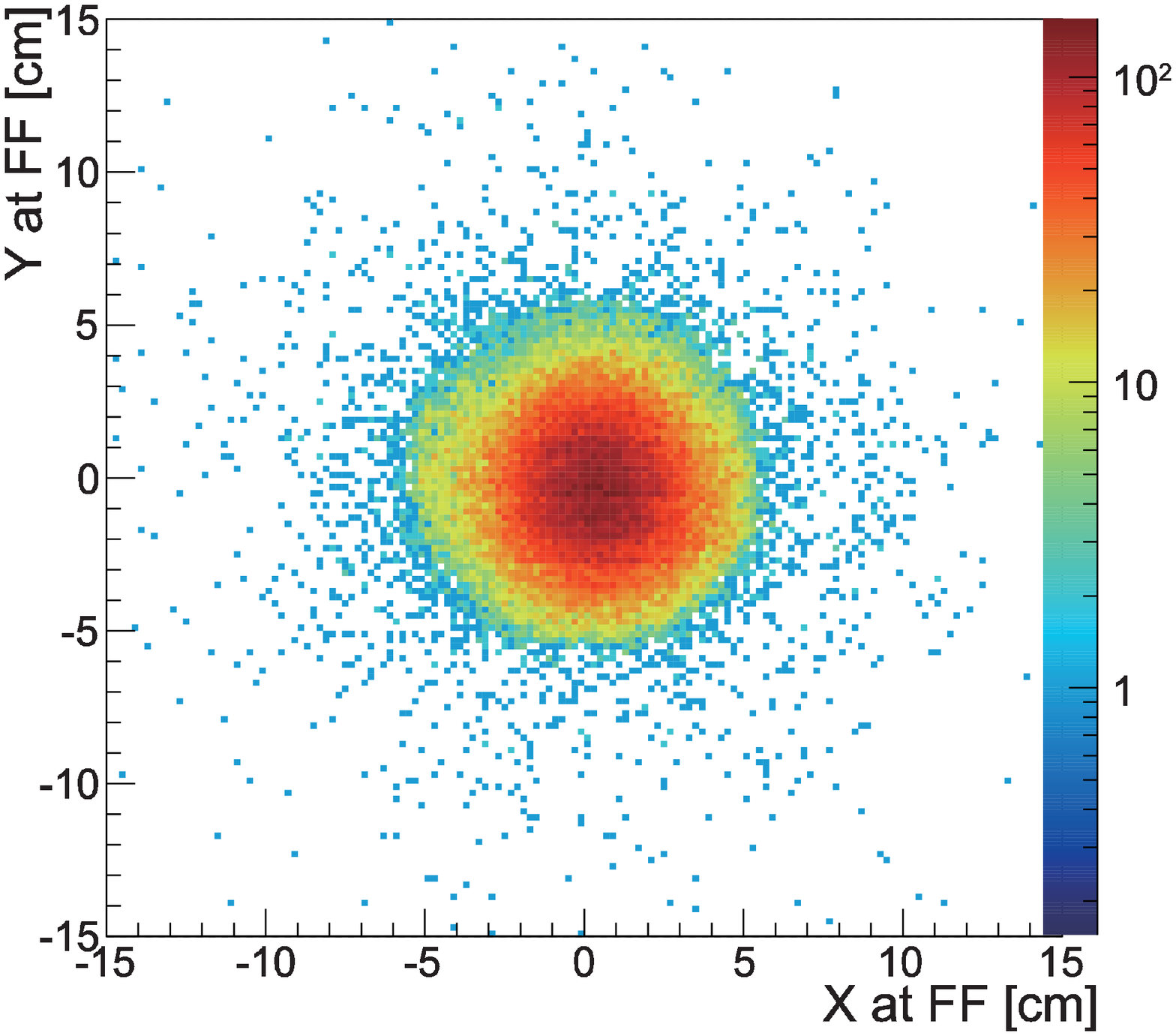}
   \caption{Beam profile of the kaons at the final focus point
   reconstructed by the BPC.}
   \label{fig:FF-prof-BPC1}
  \end{center}
\end{figure}

\begin{figure}[htbp]
  \begin{center}
   \includegraphics[width=\textwidth]{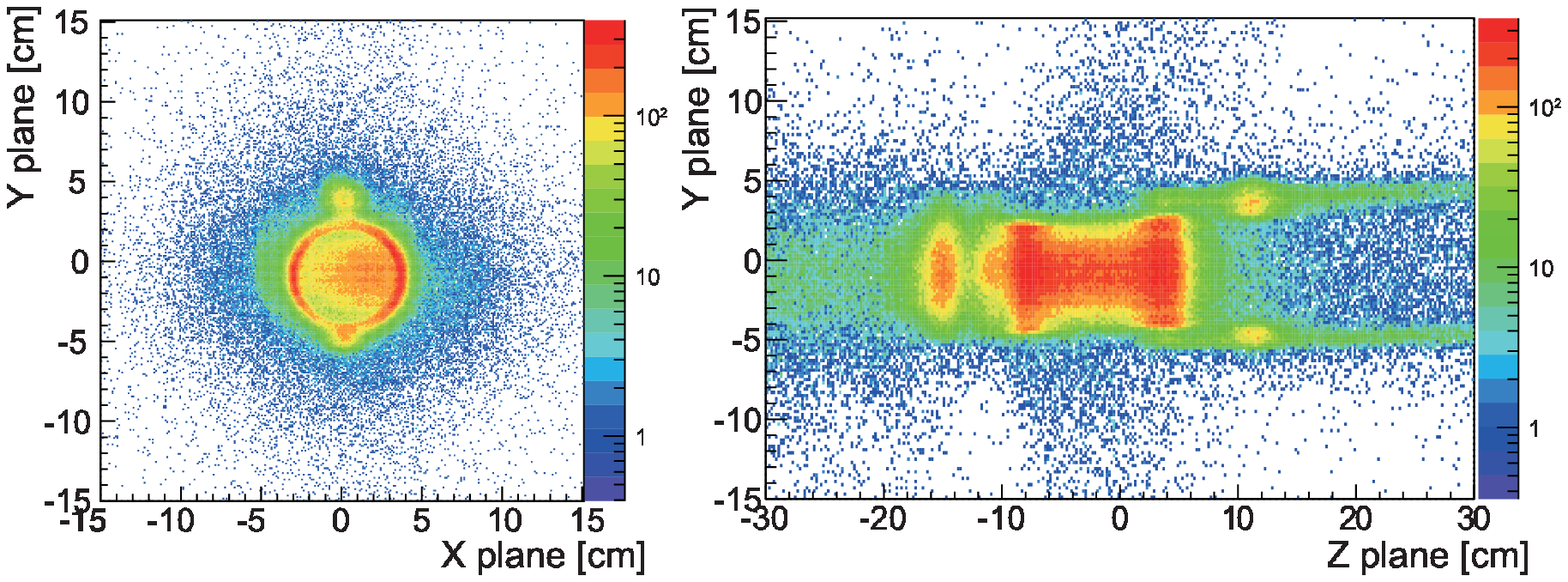}
  \end{center}
 \caption{Event vertex with the pion beam reconstructed by the CDS.
 The cross sections of perpendicular (left) and parallel (right)
 directions to the beam axis are shown.
 The liquid helium target cell and transfer pipes are clearly seen.}
 \label{fig:CDC-target}
\end{figure}

An engineering run with the full setup of the E15 experiment, i.e., the
CDS with the liquid $^3$He target system and the neutron counter, was
also conducted in June 2012.
Charged particles from the target were tracked by the CDC, and momentum
information was obtained with a magnetic field of 0.7~T provided by
the solenoid magnet.
The event vertex was obtained from trajectories of the beam and the
secondary particles that were reconstructed by the BPC and the CDC,
respectively.
The reconstructed beam profile at the final focus point is shown in
Fig.~\ref{fig:FF-prof-BPC1}, and Fig.~\ref{fig:CDC-target} demonstrates
the event-vertex reconstruction in which the liquid helium target cell
and transfer pipes are clearly seen.

\begin{figure}[htbp]
 \begin{minipage}{0.5\textwidth}
  \begin{center}
   \includegraphics[width=\columnwidth]{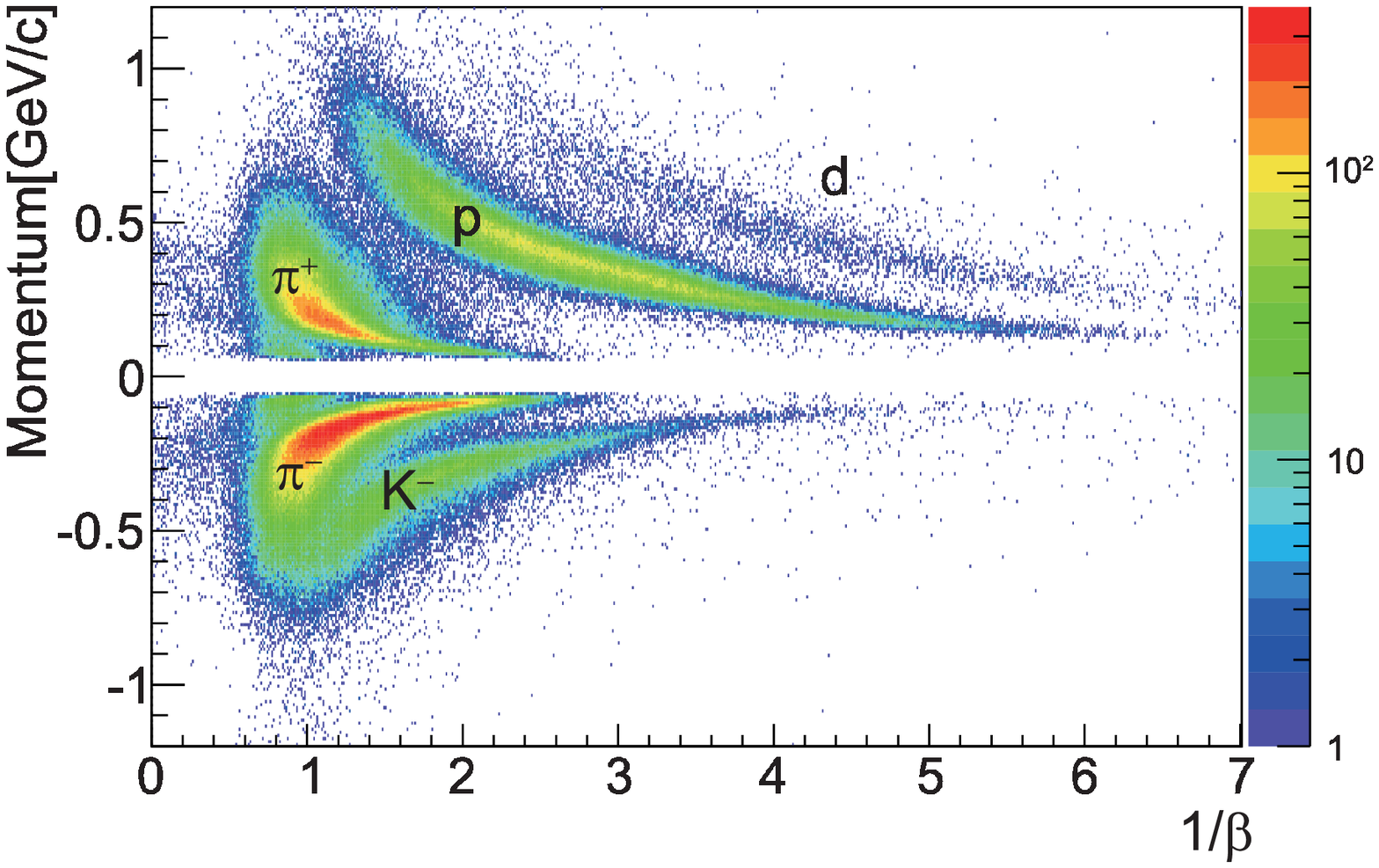}
   \caption{Distributions of the momentum versus 1/$\beta$ obtained
   by the CDS.
   }
   \label{fig:pid}
  \end{center}
 \end{minipage}
 \begin{minipage}{0.5\textwidth}
  \begin{center}
   \includegraphics[width=\columnwidth]{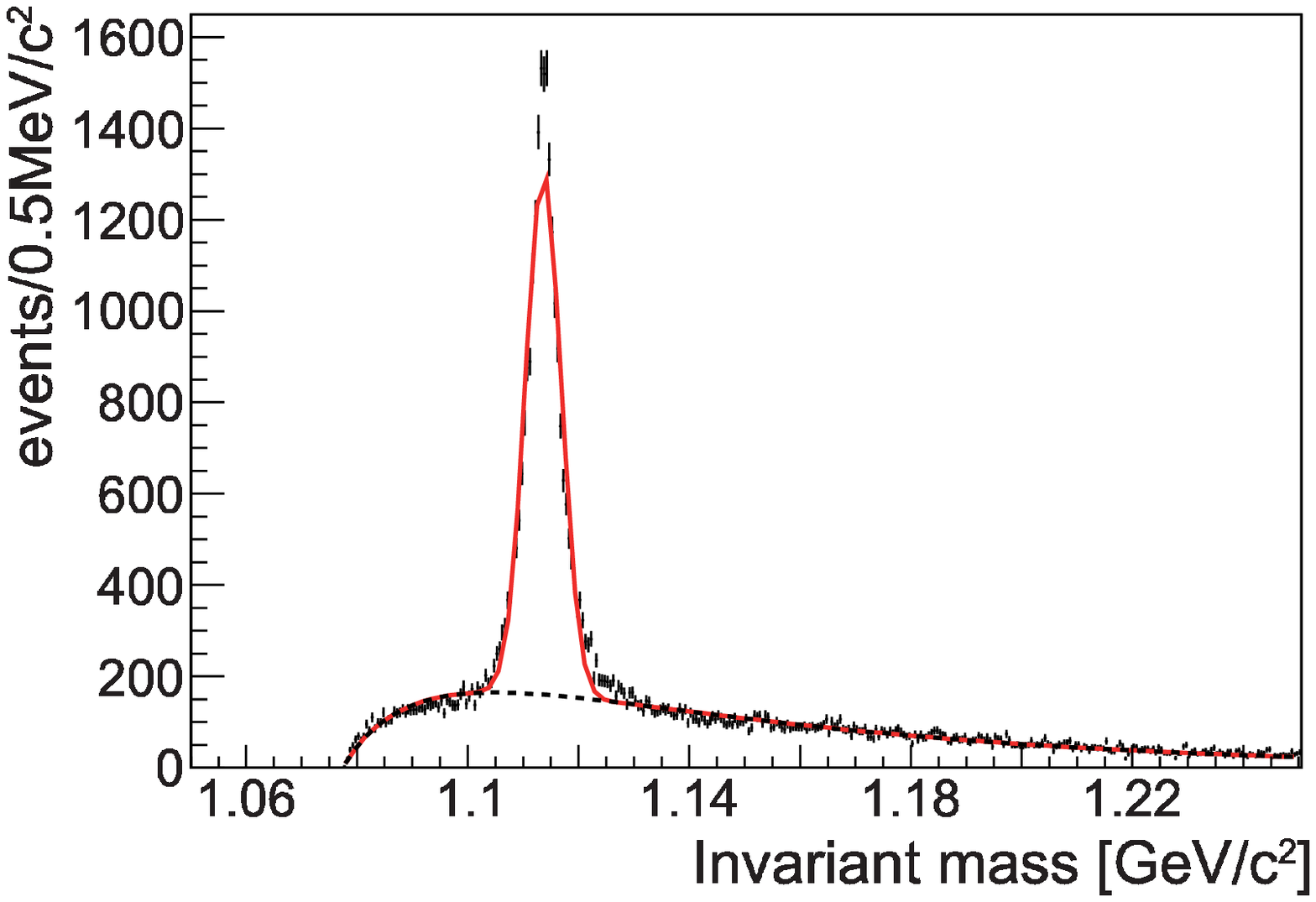}
   \caption{Invariant mass spectrum of $p \pi^-$.
   The spectrum is fitted with a Gaussian and
   a background curve.
   A displaced vertex cut of 2~cm is applied.
   }
   \label{fig:lambda}
  \end{center}
 \end{minipage}
\end{figure}

The identification of secondary charged particles was performed by
the CDH using TOF measurements together with T0.
Here the flight length was calculated from the event vertex and the
particle trajectory.
Figure~\ref{fig:pid} shows the distributions of the momentum versus
1/$\beta$.
Pions, kaons, protons, and deuterons are clearly separated.
The TOF resolution between T0 and the CDH is typically 160~ps
($\sigma$).
Using the momentum reconstruction and the particle identification,
the invariant mass of $p \pi^-$ was reconstructed as shown in
Fig.~\ref{fig:lambda}.
A clear peak of $\Lambda \to p \pi^-$ decay can be seen.
As a direct demonstration of the spectrometer performance, the mass
resolution of $\Lambda \to p \pi^-$ decay was compared to the
expectations from the detailed detector simulation.
The centroid of $\Lambda$ is obtained as 1113.6 $\pm$ 0.1~MeV/$c^2$ (known
to be 1115.7 MeV/$c^2$) with a Gaussian resolution of 3.5 $\pm$
0.1~MeV/$c^2$, whereas the expected centroid is 1113.4~MeV/$c^2$ with a
resolution of 3.5~MeV/$c^2$; the CDS performance has been reproduced by
the simulation.
The expected invariant mass resolution of the $K^-pp \to \Lambda p$
decay is evaluated to be 10 MeV/$c^2$ with the simulation, which is
sufficient to satisfy the E15 requirement of less than 20 MeV/$c^2$
($\sigma$).

\begin{figure}[htbp]
  \begin{center}
   \includegraphics[width=0.5\columnwidth]{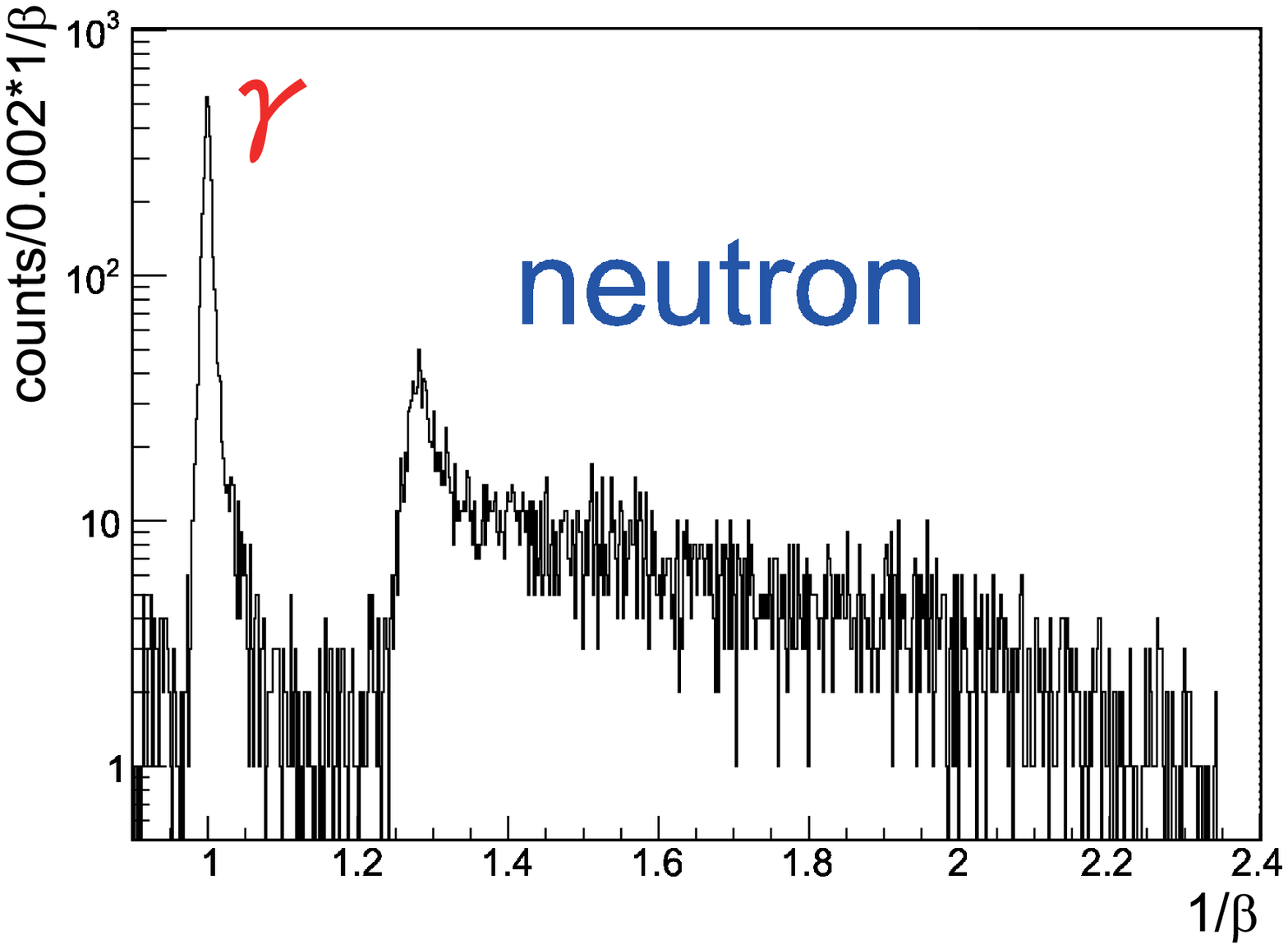}
   \caption{$1/\beta$ spectrum of the neutral particles obtained by the
   neutron counter.
   }
   \label{fig:NC-TOF}
  \end{center}
\end{figure}

Forward-going neutral particles were also successfully detected and
identified by the neutron counter in the engineering run.
The resultant $1/\beta$ spectrum of the neutral particles is shown in
Fig.~\ref{fig:NC-TOF}, in which charged particles are vetoed by the beam
veto counter and the charge veto counter.
$\gamma$ rays and neutrons are clearly separated in the spectrum.
The TOF resolution between T0 and the neutron counter is typically
150~ps ($\sigma$).
With the measured TOF resolution, the expected missing-mass resolution
for the $^3$He$(K^-,n)K^-pp$ reaction is evaluated to be 9~MeV/$c^2$
($\sigma$), which satisfies the E15 requirement of less than 10 MeV/$c^2$
($\sigma$).

\section{Summary}
A new spectrometer system was designed and constructed at the secondary
beam line K1.8BR in the hadron hall of J-PARC.
The experiments at the K1.8BR beam line aim to investigate $\bar K N$
interactions and $\bar K$-nuclear bound systems.
The E15 (kaonic-nuclei search via in-flight $^3$He$(K^-,  n$)), the E17
(kaonic-atom X-ray measurement with stopped-$K^-$), and the E31
(spectroscopic study of $\Lambda(1405)$ via in-flight $d(K^-, n$))
experiments were proposed and approved at the K1.8BR beam line.
The spectrometer consists of a high precision beam line
spectrometer, a liquid $^3$He/$^4$He/D$_2$ target system, a
cylindrical detector system that surrounds the target to detect the
decay particles from the target region, and a neutron time-of-flight
counter located $\sim$15~m downstream from the target position.
Commissioning of the beam line spectrometer and an engineering run with the
full setup of the E15 experiment were successfully performed with a 1.0
GeV/$c$ kaon beam in February and June 2012, respectively.
The results obtained show that the design goal of the spectrometer has
been achieved.
The experiments at K1.8BR are now ready, and the physics output will be
reported in the near future.

\section*{Acknowledgments}
We gratefully acknowledge all the staff members at J-PARC.
In particular, we would like to sincerely thank them for the great deal of
effort made to recover from catastrophic damage caused by the earthquake
on 11 March 2011.
This work was supported by RIKEN, KEK, RCNP, a Grant-in-Aid for
Scientific Research on Priority Areas [No. 17070005 and No. 20840047],
a Grant-in-Aid for Specially Promoted Research [No. 20002003], a
Grant-in-Aid for Young Scientists (Start-up) [No. 20028011], a
Grant-in-Aid for Scientific Research on Innovative Areas [No. 21105003],
and the Austrian Science Fund (FWF) [P20651-N20].

%

\end{document}

%% file: intro.tex
At the K1.8BR beam line, a liquid $^3$He/$^4$He and a liquid D$_2$ target will be used in the J-PARC E15/E17 and E31 experiments, respectively. 
Both target systems are combined with the CDS, which is described later. 
To have large acceptance for the secondary charged particles, an L-shaped cryostat was adopted to place the target cell at the center of the CDS.
In this section, we present an overview of both cryogenic target systems.

%% file: lhe3.tex
\subsubsection{Configuration and operational procedure} 
\begin{figure}[t]
\includegraphics[width=\columnwidth]{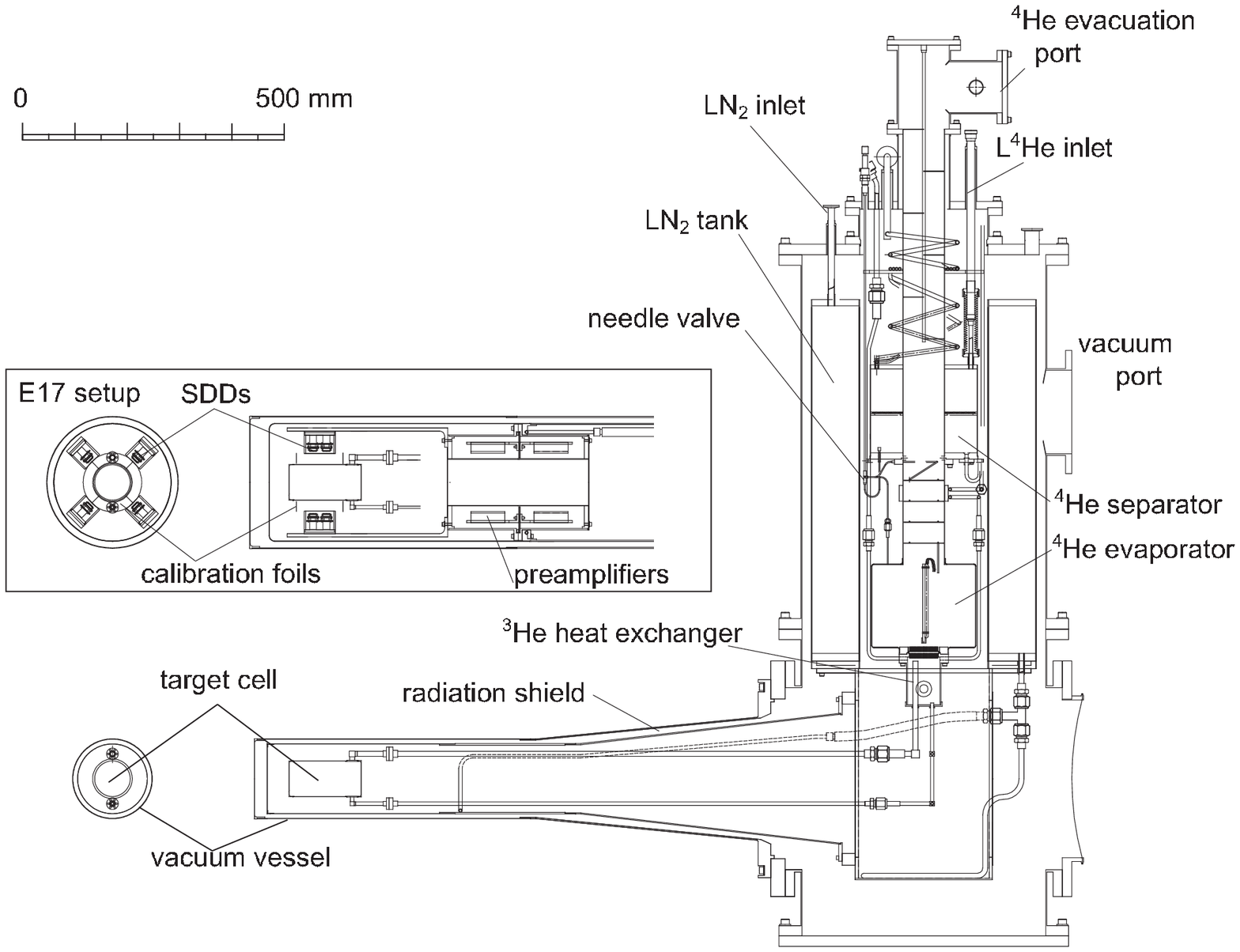}
\caption{ \label{lhe3:cryo}
Schematic drawing of the liquid $^3$He cryostat.
}
\end{figure}

A schematic drawing of the liquid $^3$He cryostat is shown in Fig. \ref{lhe3:cryo} for the case of the J-PARC E15 setup. 
The details of the $^3$He target system can be found in a separate paper \cite{Iio12}. 
The major difference between the E15 and E17 settings is the configuration around the target cell. 
To maximize the acceptance for the kaonic helium X-rays in E17,  eight silicon drift detectors (SDDs) will be installed around the target cell, as shown
 in the inset of Fig. \ref{lhe3:cryo}. 
In contrast, a time projection chamber will be installed between the target vacuum chamber and the CDS in the E15 setup.  
Thus, the diameter of the vacuum chamber is minimized as much as possible.
The major cryogenic component is divided into three sections: a $^4$He separator,  a $^4$He evaporator, and  a heat exchanger between $^3$He and $^4$He.
The target cell is connected to the bottom of the heat exchanger with two 1~m long pipes.
To reduce the radiation from room temperature components, all low-temperature parts are covered with a radiation shield  anchored to the liquid nitrogen tank (LN$_2$ tank). 

The operational concept of the E15 and E17 cryostats is essentially the same. 
The typical start-up procedure begins with liquid nitrogen cooling. 
When the $^4$He separator and the LN$_2$ tank are filled, the evaporator, the heat exchanger, and the target cell are cooled by thermal conduction and radiation.
After the pre-cooling, the liquid nitrogen in the separator is purged, and the liquid helium is transferred from a 1000 $\ell$ Dewar (not shown in the figure) to the separator by a transfer tube. 
The liquid flow is controlled by the pressure inside the separator evacuated by a dry pump. Liquid $^4$He inside the separator is fed to the $^4$He evaporator though a needle valve.
The vapor pressure in the evaporator is reduced by a rotary pump with a pumping speed of 120 m$^3$/h, resulting in a heat-removal capability of 2.5 W at 2 K. 
The temperature inside the evaporator is controlled within a range of  1.3 to 2.0 K. This range is mainly determined by the flow rate from the separator to the evaporator. The lowest temperature is achieved with no flow from the separator because the liquid temperature in the separator is fairly high (4.2 K). For liquefaction of $^3$He, the heat exchanger between liquid $^4$He and gaseous $^3$He is positioned below the evaporator. The top part of the heat exchanger, where the liquid $^4$He in the evaporator is in direct contact, has a specially designed fin structure with both a width and a pitch of 0.5 mm. 

A gas-tight handling system (leak rate less than 10$^{-10}$ Pa$\cdot$m$^3$/sec)
 has been constructed to store, transfer, and recover the scarce $^3$He gas.
The total amount of 400 $\ell$ of gaseous $^3$He is stored at pressures of less than an atmosphere at room temperature in two 200 $\ell$ tanks. During the cooling stage, these gas tanks are connected to the heat exchanger through the gas handling system.  
 By an effective heat contact inside the heat exchanger, gaseous $^3$He is liquefied, and the liquid $^3$He flows to the target cell (6.8~cm in diameter and 13.7~cm in length) through the lower pipe.
 In the last stage of cooling, most of the $^3$He gas is liquefied inside the target cell and the heat exchanger. 

In the L-shaped cryostat, the heat load on the target cell must be transferred effectively to the heat exchanger where the cooling power exists, otherwise boiling in the target cell occurs.
To accomplish this, we applied the {\it siphon method} as described in Ref. \cite{Iio12}, which uses convection of the liquid $^3$He.
The liquid $^3$He warmed by the heat load inside the target cell returns to the heat exchanger through an upper pipe. In the heat exchanger, $^3$He is cooled again and fed to the target cell through the lower pipe. This makes the heat transfer between the target cell and the heat exchanger possible.  

For long-term operation, it is essential to reduce the total amount of $^4$He consumed. This is because exchanging the $^4$He Dewar causes significant experimental dead time. 
To minimize the $^4$He consumption, we adopted {\it one-shot} operation. 
This operation consists of two modes:
(I) the evaporator is filled up with liquid $^4$He supplied from the separator. (II) The $^4$He supply is stopped until the evaporator becomes empty. 
The operational procedure consists of a repetition of these two methods, and 
this reduces the total liquid $^4$He consumption due to the minimization of the transfer loss to the cryostat. 
The operational performance of the target system is described in the following subsection.

\subsubsection{Operation and performance} 
Along with the operational procedure previously described, cooling tests were performed. 
After the $^4$He transfer, it took about 2-3 hours to liquefy  the $^3$He gas in the heat exchanger, achieving thermal equilibrium within 6 hours and a temperature of 1.30 $\pm$ 0.01 K in the target cell without flow from the separator (mode (II) in the {\it one-shot} operation).
The liquid $^3$He density at this temperature is 0.0812~g/cm$^3$, corresponding to a thickness of 1.11~g/cm$^2$.
The density fluctuation due to the temperature instability is less than 0.1\%.
The temperature differences between the evaporator, the heat exchanger, and the target cell are less than 0.01 K. This means that the heat transfer by the {\it siphon method} is working well.
Furthermore, the pressure inside the heat exchanger was identical to the vapor pressure of liquid $^3$He at the corresponding temperature. Taking into consideration the remaining pressure inside the tanks, a total amount of 380 $\ell$ was condensed, giving evidence that sufficient $^3$He gas is liquefied to fill the target.

From the reduction rate of the liquid $^4$He in the evaporator, 
the heat load of the low-temperature region was estimated to be 0.21 W with the E15 setting.
On the E17 setting, the heat load was expected to increase due to the radiation from the SDDs to the target. 
It was measured to be 0.39 W with the actual E17 setting, and both of these are acceptably small for long-term operation.
The  operational results of the cryostat with the E15 setting are tabulated below.

\begin{table}[h] 
\begin{center}
\caption{\label{operation}Operational results.}
\begin{tabular}{lrc}\hline\hline
Vacuum level                      & [mbar]                 &$ < 10^{-6}$ \\ 
Leak rate of the $^3$He system    & [Pa$\cdot$m$^3$/sec] &$< 10^{-10}$ \\ 
Temperature in the target cell    &[K]       & 1.3        \\
Vapor pressure in the target            &[mbar]    & 33         \\
Heat load to low-temperature part   &[W]       & 0.21       \\ 
Liquid $^4$He consumption          &($\ell$/day)      & 50         \\ 
 \hline \hline  
\end{tabular}
\end{center}
\end{table}
Finally, we note that this cryostat can be utilized as a liquid $^4$He target system by liquefying gaseous $^4$He instead of $^3$He. The operational procedure and the performance of the liquid $^4$He target are the same as those of $^3$He.
The density of the liquid $^4$He is 0.145~g/cm$^3$ at 1.3~K with a stability of better than 0.1\%, and the thickness is 1.99 g/cm$^2$.


%% file: d2.tex
\begin{figure}[t]
\includegraphics[width=\columnwidth]{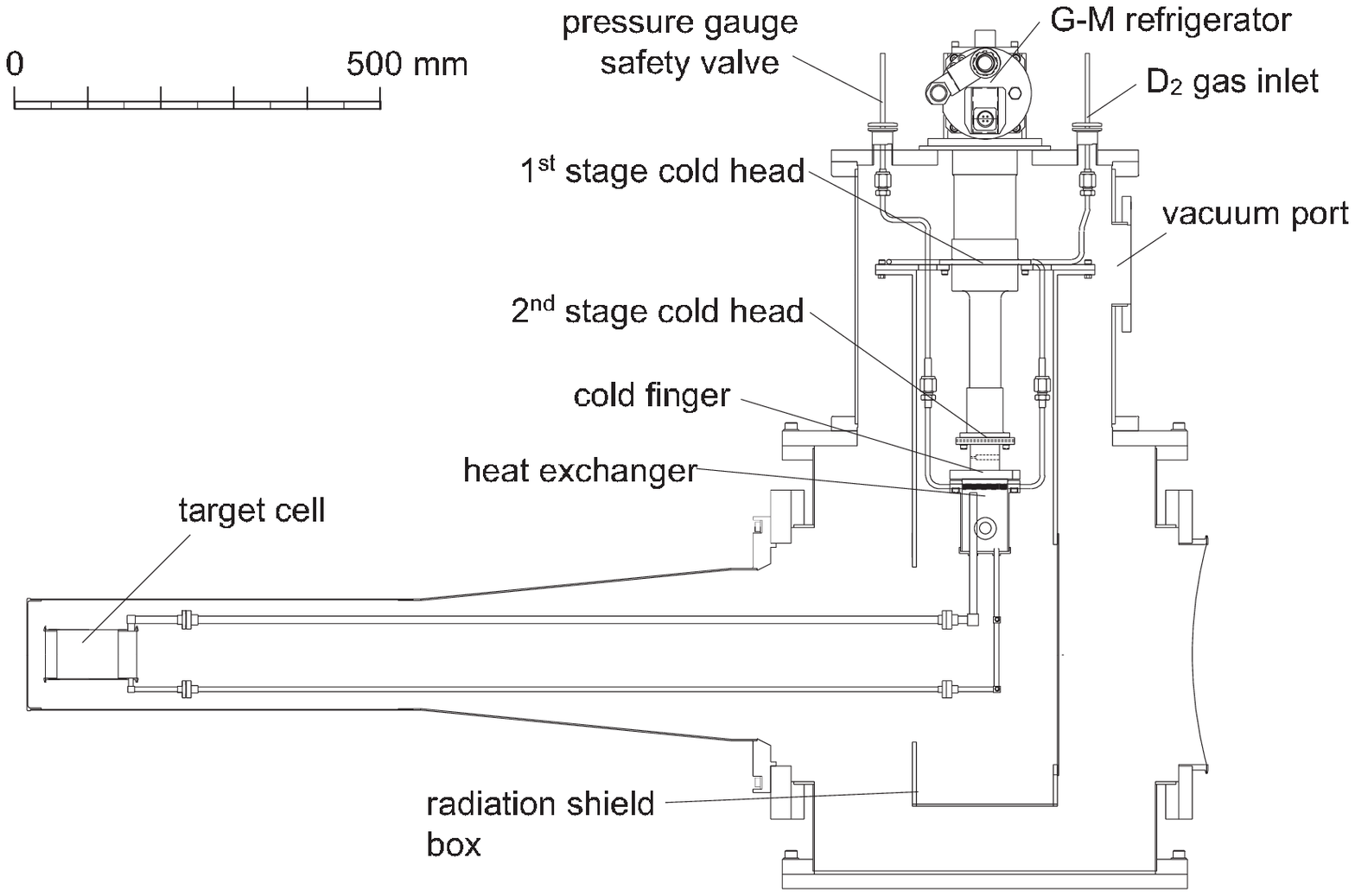}
\caption{ \label{d2:cryo}
Schematic drawing of the liquid D$_2$ cryostat.}
\end{figure}

For a spectroscopic study of $\Lambda$(1405) by the $d(K^-,n)$ reaction (J-PARC E31), we have been developing a liquid D$_2$ target system.
A side-view of the cryostat is shown in Fig. \ref{d2:cryo}. Since we measure the decay products of $\Lambda$(1405), the target cell, whose size is 6.8~cm in diameter and 12.5~cm in length, is isolated at the center of the CDS in the same way as the liquid $^3$He cryostat. 
The major difference from the  liquid $^3$He target is that this cryostat is coolant-free. The key component of the system is a two-stage Gifford-McMahon (G-M) refrigerator (Sumitomo Heavy Industries, Ltd.,  RDK-145D and CSA-71A) built into the cryostat.
 The cooling power at the first and second stages is 35 W at 50 K and 1.5 W at 4.2 K, respectively. 
As it has been for the $^3$He target, a gas handling system has been constructed for the D$_2$ target. 
The 1000 $\ell$ of gaseous D$_2$ is stored in a tank at 2 bar at room temperature.  
To avoid contamination, the amount of D$_2$ gas was chosen to maintain positive pressure inside the gas system even after liquefaction in the target.

The D$_2$ gas is fed into the  cryostat through the top flange.
For pre-cooling, the inlet pipe for the D$_2$ gas is anchored to a copper plate attached to the first-stage cold head of the G-M refrigerator. 
Another inlet pipe is directly connected through the top flange to the heat exchanger. This is used to measure the D$_2$ pressure inside the heat exchanger. 
Since this pipe has a larger conductance, a safety valve that prevents a sudden pressure rise is also connected to it.
The D$_2$ gas is cooled in the heat exchanger where the second stage of the G-M refrigerator is thermally contacted. 
The structure of the heat exchanger is similar to that of the $^3$He system \cite{Iio12}.
The main difficulty in the operation of the system is the precise control of the temperature in the heat exchanger. 
In the liquid D$_2$ target system, the {\it siphon method}, described previously, is also adopted. 
For effective heat transfer between the target cell and the heat exchanger, D$_2$ must be kept in a liquid state by controlling the temperature to avoid blocking of pipes by solid D$_2$. 
The temperature range of liquid D$_2$ is 18.7-23.8 K at 1 bar, thus the
temperature should be kept around 20 K within acceptable limits.
The thickness of D$_2$ along the beam is 2.13~g/cm$^2$ with a density of 0.17~g/cm$^3$ at 20 K.
Since the cooling power of the second stage of the G-M refrigerator is larger than the heat load on the low-temperature parts, we have installed a heater near the cold finger to compensate the heat load. 
The current in the heater is controlled by a proportional-integral-derivative (PID) algorithm 
 with an input of the temperature of the heat exchanger.

So far, 
we have installed the G-M refrigerator and heat exchanger in the vertical part of the cryostat. Using hydrogen gas for convenience, 
preliminary tests were performed to confirm the temperature stability.  
In these tests, the target cell was directly connected to the heat exchanger.
The result showed that the temperature of the heat exchanger is controlled within 0.2 K, which is sufficiently stable for the operation. 
By the end of 2012, cooling tests will be performed on the final setup with D$_2$ gas.

%% file: sdd.tex
The goal of the E17 experiment is to achieve a precision measurement of both kaonic-$^3$He and -$^4$He atoms with an unprecedented accuracy of 1 eV. To achieve this precision, we employ silicon drift detectors (SDDs) with a large active area. 

The concept of an SDD was originally introduced by E. Gatti and P. Rehak in 1984 \cite{Gatt84}.
The electric field parallel to the surface of the detector is generated by ring electrodes biased gradually. 
Electrons created by incoming X-ray absorption drift toward a collection anode placed at the center of the detector.
The distinguishing feature of SDDs is the extremely small anode size,
which results in the low capacitance of the detector.
It is also independent of the detector active area, thus a large active area size of 100 mm$^2$ becomes possible with low capacitance.
To take advantage of the low output capacitance, an FET for the first-stage amplification is directly integrated on the detector chip. It is connected to the anode with a short metal strip to minimize the stray capacitance and microphonic noise. 
A typical energy resolution of 150~eV is obtained at 6~keV with sufficient noise reduction. The time resolution is typically sub-micro seconds below 200 K, which is mainly determined by the drift-time distribution of electrons in silicon. 
In recent years, several types of SDDs with a large active area have been developed. For X-ray spectroscopy of kaonic atoms, SDDs were used in the KEK-PS E570 experiment in KEK\cite{E570} and the SIDDHARTA experiment in LNF\cite{SIDDHARTA}.

In the J-PARC E17 experiment, 8 SDDs and reset-type preamplifiers developed by KETEK\footnote{KETEK GmbH, Vitus-SDD without window and collimator} were adopted.
Each detector has a thickness of 0.45 mm and an active area of 100 mm$^2$.
They are mounted around the liquid helium target as illustrated in the inset of Fig. \ref{lhe3:cryo}. 
The acceptance for both kaonic $^3$He and $^4$He $L_{\alpha}$ X-rays is approximately 1 \% with 8 SDDs, taking into account the attenuation inside the helium target, the target cell, etc. 
The preamplifiers are also installed in vacuum to minimize the cable length between SDDs and preamplifiers. Output signals from the preamplifiers are connected to a CAEN N568b shaping amplifier. Semi-Gaussian outputs are provided with different shaping times of 0.5 and 3.0 $\mu$s.  Output signals with 0.5 $\mu$s shaping time are used for the timing information of the SDDs in coincidence with an incoming $K^-$. Signals with 3.0 $\mu$s shaping time are recorded with two types of peak-hold ADCs, a TKO peak-hold ADC and a VME CAEN V785. For the purposes of pileup rejection, the line-shapes of the signals are also recorded by a flash ADC (SIS3301, 14 bit, 105~MHz).

To reduce the heat loads on the helium target, the SDDs must be  operated at low temperature. 
We performed basic studies on the temperature dependence of the energy and time resolutions
,  and optimized the operational temperature of the SDDs to 130 K. 
Heat load from the preamplifiers is minimized by covering them with heat shields cooled down to 77 K.
All 8 SDDs were operated successfully inside the cryostat during cooling of the  liquid $^4$He target. 

\begin{figure}[t]
\begin{center}
\includegraphics[width=0.5\columnwidth]{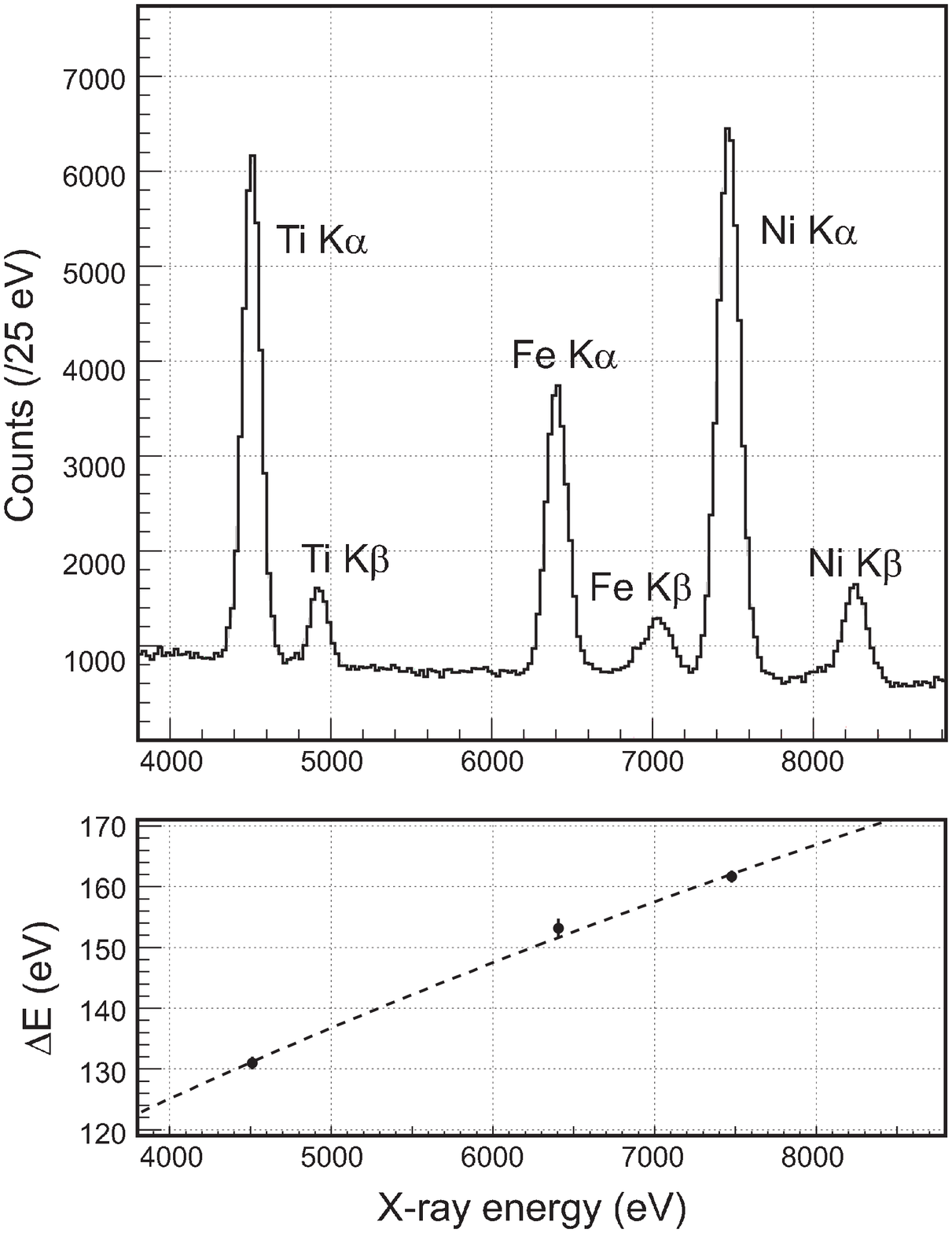}
\caption{\label{sdd:spectrum} A spectrum with fluorescence x-rays induced by the beam at  K1.8BR (top) and the energy dependence of the energy resolution (bottom). The dotted line represents an empirical formula described in the text. 
}
\end{center}
\end{figure}

In November 2010, we performed commissioning of the SDDs with the secondary beam at the K1.8BR beam line.  The spectrum obtained with 8 SDDs is shown in Fig. \ref{sdd:spectrum}. 
Absolute energy calibration was performed with $K_{\alpha}$ fluorescence X-rays from titanium (4.5 keV) and nickel (7.5 keV) foils induced by the beam particles.
Furthermore we installed an iron foil at the target position to have fluorescence X-rays at an energy around 6.5 keV. 
As shown in the lower panel of Fig. \ref{sdd:spectrum}, we obtained the energy resolution (FWHM) for three different energies.
The dotted line in the figure is an empirical formula, $2.35 \omega \sqrt{W_N^2 + FE/\omega}$, where $E$, $\omega$, $W_N$,  and $F$ are incident X-ray energy,  electron-hole pair creation energy, noise constant, and Fano factor, respectively. 
The energy dependence is well understood by the known formula.
The obtained values are $F = 0.14 \pm 0.01$ and $W_N = 6.7 \pm 0.8$ with a constant of $\omega = 3.81$ eV.
We achieved an energy resolution of 150 eV FWHM at an energy for kaonic $L_\alpha$ lines in $^3$He and $^4$He, which is better than the 185~eV of KEK-PS E570.
This energy resolution is sufficient to achieve a precision of 1 eV with the expected statistics of kaonic $L_\alpha$ for 5,000 events.
The signal to noise ratio for the calibration spectrum in the commissioning  is also higher than that of E570 by a factor of 3.  
As a result of the beam commissioning, good performance of the SDDs was achieved under realistic conditions.


